\documentclass[reprint,amsmath,amssymb,aps,prd]{revtex4-2}

\usepackage{newtxtext,newtxmath}

\usepackage[T1]{fontenc}

\DeclareRobustCommand{\VAN}[3]{#2}
\let\VANthebibliography\thebibliography
\def\thebibliography{\DeclareRobustCommand{\VAN}[3]{##3}\VANthebibliography}

\DeclareUnicodeCharacter{2212}{-}
\newcommand{\jcap}{Journal of Cosmology and Astroparticle Physics}

\usepackage{graphicx}	
\usepackage{amsmath}	

\begin{document}
\title{The Preference for Evolving Dark Energy from Cosmological Distance Measurements and Possible Signatures in the Growth Rate of Perturbations}

\author{Ryan E.\ Keeley}
  \email{rkeeley@ucmerced.edu}
 \affiliation{University of California, Merced, 5200 N Lake Road, Merced, CA 95341, USA}

\author{Kevork N.\ Abazajian}
\affiliation{Department of Physics and Astronomy, University of California - Irvine, Irvine, California 92697, USA}

\author{Manoj Kaplinghat}
\affiliation{Department of Physics and Astronomy, University of California - Irvine, Irvine, California 92697, USA}

\author{Arman Shafieloo}
\affiliation{Korea Astronomy and Space Science Institute, 776, Daedeokdae-ro, Yuseong-gu, Daejeon
34055, Republic of Korea}

\label{firstpage}

\begin{abstract}
In this study, we use a flexible parametrization of the equation of state of dark energy to explore its possible evolution with datasets from the Dark Energy Spectroscopic Instrument (DESI), Planck cosmic microwave background (CMB), and either the 5-year Dark Energy Survey (DES) or the Pantheon+ (PP) supernova (SN) compilation.  
This parametrization, called transitional dark energy (TDE), allows for rapid changes in the equation of state but also changes like that in the Chevallier-Polarski-Linder (CPL) $w_0$–$w_a$ parametrization. 
We find a 3.8$\sigma$ preference for evolving dark energy over $\Lambda$CDM with the DES SN dataset and a weaker 2.4$\sigma$ preference when using the PP dataset. 
This corroborates the finding of the DESI Collaboration, who found that their baryon acoustic oscillation (BAO) data preferred  evolving dark energy when fit with the CPL parametrization of the equation of state. 
Our analysis reveals no significant outliers in the DESI data around the TDE best-fit, while the data is asymmetrically distributed around the $\Lambda$CDM best-fit model such that the measured distances are on average smaller. 
The DESI and SN data both prefer an expansion history that implies a higher dark energy density around $z=0.5$ than in the Planck-$\Lambda$CDM model, with the inferred equation of state being greater than $-1$ around $z=0$ and close to or below $-1$ at $z>0.5$. 
We show that when the expansion rate is greater than that in the Planck-$\Lambda$CDM model (around $z = 0.5$), the growth rate calculated assuming General Relativity is suppressed relative to the Planck-$\Lambda$CDM model, and it rebounds as the expansion rate differences between the models become smaller closer to the present time. 
We derive an approximate analytic expression for the growth rate differences to understand this behavior quantitatively. 
The resulting flattening of the $f\sigma_8(z)$ curve compared to the $\Lambda$CDM model could be an independent signature of the temporal evolution of dark energy.   
\end{abstract}

\maketitle



\section{\label{sec:intro}Introduction}
The $\Lambda$CDM model is the current concordance model of cosmology. 
The $\Lambda$ component, representing a constant dark energy density, makes up approximately 70\% of the Universe, while cold dark matter (CDM) and baryons account for the rest. 
This model explains various observations of the Universe, including the present accelerated expansion ~\cite{SupernovaSearchTeam:1998fmf,SupernovaCosmologyProject:1997czu,Frieman:2008sn,2013PhR...530...87W,Pan-STARRS1:2017jku}, the large-scale structure ~\cite{SDSS:2005xqv}, and the cosmic microwave background (CMB)~\cite{Bennett:1996ce}. 

Current datasets that constrain $\Lambda$CDM include Type Ia supernova (SN) compilations from Pantheon+ (PP)~\cite{Brout:2022vxf}, 5-year Dark Energy Survey (DES)~\cite{DES:2024tys} and Union3~\cite{Rubin:2023ovl}, baryon acoustic oscillation (BAO) datasets from DR16 SDSS/eBOSS~\cite{eBOSS:2020yzd} and Dark Energy Spectroscopic Instrument (DESI)~\cite{DESI:2024mwx}, as well as CMB measurements from Planck~\cite{Planck:2018vyg}, Atacama Cosmology Telescope (ACT)~\cite{ACT:2023dou,ACT:2023kun}, and the South Pole Telescope (SPT)~\cite{SPT:2023jql}. 
In this precision era of cosmology, the goal is to test the $\Lambda$CDM model.
Recent results from DESI have detected a 2-4$\sigma$ preference for an evolving dark energy model over the concordance $\Lambda$CDM models in which the dark energy density is constant in time~\cite{DESI:2024mwx,DESI:2024kob,DESI:2024aqx}.  
The DESI results were based on fitting a CPL functional form for the equation of state with two parameters $w_0\,\&\,w_a$ (present day equation of state and derivative with respect to the scale factor of the Universe)~\cite{Chevallier:2000qy,Linder:2002et}.

One motivation for an evolving dark energy model was its possible solution to the $H_0$ and $S_8$ tensions.
The $H_0$ tension is a mismatch in the value of the present Hubble rate ($H_0$) as directly measured by the Cepheid calibration of SN distances~\cite{Riess:2016jrr,Riess:2019cxk,Riess:2020fzl}, and the value as inferred from the CMB assuming a $\Lambda$CDM model~\cite{Planck:2018vyg}.  
This mismatch has reached a statistical significance of over $5\sigma$~\cite{Riess:2021jrx}.  
The $S_8$ tension is similar, where there is a mismatch between the $S_8$ parameter (which measures the amplitude of matter clustering) inferred from Planck CMB measurements, assuming the $\Lambda$CDM model, and that measured with low-redshift weak lensing datasets, though the preference is less significant than the $H_0$ tension~\cite{KiDS:2020suj,DES:2021wwk,Kilo-DegreeSurvey:2023gfr}. 
Here, $S_8 \equiv \sigma_8 (\Omega_m/0.3)^{0.5}$, where $\sigma_8$ is the amplitude of the matter power spectrum at the $8\,\mathrm{Mpc}/h$ scale and $h\equiv H_0/(100\,\mathrm{km\,s^{-1}\,Mpc^{-1}})$.
With new data, it has become evident that changes to low-redshift cosmology cannot solve the $H_0$ tension~\cite{Keeley:2022ojz}, but there is significant room for evolution in the dark energy density with time.

In this work, we fit a more flexible parametrization of the dark energy equation of state, called transitional dark energy (TDE), to the Planck CMB, DESI BAO, and the 5-year DES or the PP SN datasets to both constrain the expansion history and predict the corresponding growth history, assuming General Relativity. The TDE parametrization allows for expansion histories similar to those in the CPL model and also sharper changes in the temporal evolution of the dark energy density. Using the greater freedom allowed by the TDE parametrization, we can test whether the inference for the deviation from the cosmological constant is an artifact of how the dark energy equation of state is parameterized or prior-driven, as has been suggested based on the observation that $w(z)$ is very close to $-1$ with the CPL functional form at the redshifts where SN and BAO datasets provide the strongest constraints ~\cite{Cortes:2024lgw}. 

In the present work, we will assume that the distance data are faithfully constraining the expansion history and that systematic effects are subdominant. We will further assume that the only change from the $\Lambda$CDM model is through a flexible parametrization of the equation of state of dark energy $w(z)$ that relegates the effect of dark energy to late times. While the functional form assumed for $w(z)$ is fairly general, the choice still restricts the dark energy density to be positive and it may not capture the behavior of, say, rolling scalar fields with a negative cosmological constant, or interaction between dark energy and other components of the energy density~\cite{Visinelli:2019qqu,DiValentino:2020kpf,Calderon:2020hoc}. The results we obtain for $w(z)$ can be mapped onto canonical models of rolling scalar fields, but we do not attempt that here. 
Other works have attempted to map the expansion histories preferred by DESI's fit using the CPL parametrization onto quintessence models~\cite{Shlivko:2024llw}. Similar works have found that different parametrizations of the dark energy equation of state can account the dark energy phenomenology that fits the DESI data well~\cite{Wolf:2025jlc}.

\section{Data}\label{sec:Data}

The DESI BAO dataset~\cite{DESI:2024mwx} measures the two-point correlation function of galaxies in six tracers, in seven redshift bins from $0.3<z<1.49$.  These correlation functions contain a `BAO feature' which is an excess of clustering at the drag scale $r_d$ and this BAO feature can be used as a standard ruler to measure the expansion rate of the Universe. Since the true distances to these galaxies are unknown, a fiducial model has to be assumed in order to convert angles and redshifts into 3-dimensional positions. Differences between the true cosmology and the fiducial will show up as a shifts in the Alcock-Paczynski parameters~\cite{Alcock:1979mp}.  Thus, the DESI likelihoods are written in terms of either the angle-averaged volume distance $D_V(z)/r_d$, or if both the line-of-sight and transverse modes can be separated, the angular diameter distance $D_M(z)/r_d$, and Hubble distance $D_H(z)/r_d = c/H(z)/r_d$, all relative to the drag scale.  We refer to this dataset as `BAO' throughout this work.

DESI has also released a redshift-space distortion (RSD) dataset, which measures the growth rate of structure in the Universe in terms of the quantity $f\sigma_8(z)$, where $\sigma_8(z)$ is the amplitude of the matter power spectrum at scales of $8h^{-1}$ Mpc at different redshifts and $f(z)$ is the growth rate, $f = -d \log D(z) / d\log (1+z)$, where $D(z)$ is the growth function~\cite{DESI:2024hhd,DESI:2024jis}.

We also include the Type Ia SN likelihood from the full 5-year DES SN dataset~\cite{DES:2024tys}, as well as the PP dataset~\cite{Brout:2022vxf}. We include both of these two separate cases to check how the preference for evolving dark energy depends on the choice of SN dataset. Type Ia SN are useful cosmological probes since they are empirically determined to be standardizable candles. In other words, up to calibration parameters that are either internally constrained (parameters that determine how much the distance modulus shifts as a function of the observed lightcurve parameters such as the stretch and color), or externally calibrated (the intrinsic luminosity of SNe $M$ which is determined by the Cepheid distance ladder), observing the brightness of a SN allows one to infer the luminosity distance to the SN.  By observing the host galaxy of a SN, and thus measuring a redshift, one can constrain the expansion rate of the Universe through the luminosity-distance versus redshift relation---that is, up to an overall factor.  
However, since the SNe intrinsic luminosity $M$ is degenerate with $H_0$, SNe only constrain the shape of the Universe's expansion history. Whenever we include either the DES or PP SN dataset, we allow $M$ to vary as a free parameter. 

The publicly released 5-year DES SN dataset includes 1635 SN in the redshift range $0.10<z<1.13$, which includes a 5-fold increase in the number of high-redshift ($z>0.5$) SNe compared to other compilations~\cite{DES:2024tys}. We refer to this SN dataset as `DES'.  
Similarly, the PP SN dataset includes 1701 lightcurves of 1550 individual SNe. The cosmological analysis~\cite{Brout:2022vxf} represents an increased sample size and redshift coverage compared to the original Pantheon dataset~\cite{Pan-STARRS1:2017jku}. We refer to this SN dataset as `PP'.

The Planck satellite ~\cite{Planck:2018vyg} measures the anisotropies in the temperature and polarization of the CMB.  We use the `TT', `TE' and `EE' parts of the Planck 2018 dataset. We refer to this dataset as `CMB' throughout this work. To build intuition for how the Planck-CMB dataset constrains the TDE parametrization, which modifies the expansion history at low redshift, remember that the CMB primarily constrains the low-redshift expansion history via geometric degeneracies~\cite{Keeley:2019esp}. For instance, changing $H_0$, or any modification in the expansion history that changes the angular acoustic scale $\theta_*$, induces a phase-shift in the CMB's acoustic peaks~\cite{Keeley:2020rmo}. The information relevant for TDE, or any other parametrization of the dark energy equation of state, contained in the CMB can be summarized as constraints on the Hubble parameter at an angular diameter distance to the surface of last scattering $H(z_*)$ and  $D_A(z_*)$.

\section{Parametrization of the Equation of State}\label{sec:TDE}

The DESI collaboration showed that a CPL parametrization of the equation of state with a phantom crossing was preferred over $\Lambda$CDM when fitting their BAO dataset, in combination with the Planck CMB and the DES, PP or Union3 SN datsets~\cite{DESI:2024mwx}. Since we do not have satisfactory models for this kind of temporal change in the equation of state, it is worth considering if the evolution is somehow a result of the CPL parametrization and how would allowing for more flexible functions of time impact this preference.

The TDE parametrization has the following functional form for the dark energy equation of state~\cite{Keeley:2019esp}:
\begin{equation}
    w(z) = \left[(w_0 + w_1) + (w_1-w_0)\tanh\left((z-z_T)/\Delta_z\right)\right] / 2\,,
\end{equation}
where $w_0$ is the value of the equation of state below the transition redshift, $z_T$, and $w_1$ is the value of the equation of state above the transition redshift. 
The sole purpose of this parametrization is to allow for both gradual and rapid changes in $w(z)$ and give the data much more freedom in picking out the preferred expansion histories. 
The parameter $\Delta_z$ is the width of the transition and it controls how rapid or gradual the change in $w(z)$ is. 
The evolution of the dark energy density is related to $w(z)$ as follows, 
\begin{equation}
\rho_{\rm DE}(z) = \rho_0 \exp\left( 3 \int_0^z dz' \frac{1 + w(z')}{1+z'} \right)\,.
\end{equation}
For convenience, we scale the fractional dark energy density ($\Omega_{\rm DE}$) by $h^2$, as is done for the matter density ($\omega_m$), which is the sum of the dark matter and baryon densities, and define 
\begin{equation}
    \omega_{\rm DE}(z) = \Omega_{\rm DE}(z)h^2 = \frac{8 \pi G}{3}\frac{\rho_{\rm DE}(z)}{(100\,\mathrm{km\,s^{-1}\,Mpc^{-1}})^2}\,. 
\end{equation}
This allows us to write the Friedmann equation simply as,
\begin{equation}
    h^2(z) = \omega_m (1+z)^3 + \omega_{\rm DE}(z) + \omega_{\rm rad}(z)\, ,
\end{equation}
where $h(z)$ is non-dimensional Hubble parameter $H(z)/(100 \,\mathrm{km\,s^{-1}\,Mpc^{-1}})$.  We take the radiation density, $\omega_{\rm rad}(z)$, to be the same as in Ref.~\cite{Planck:2018vyg}, including how massive neutrinos are treated.

We impose a prior on the parameters $z_t$ and $\Delta_z$ such that they are allowed to vary between 0 and 10.  
Allowing $z_t$ to be negative would indicate a transition in the dark energy equation of state would happen in the future.
While such a transition could happen in the future, the data is not sensitive to allowing $z_t$ to have negative values. 
For example, in the case of $\Delta_z$ being small compared to the absolute value of $z_t$, i.e. a sharp transition, a transition in the future would just yield a constant $w(z) = w_1$ at $z>0$ in this parametrization. 
Conversely, in the case of a broad transition, $\Delta_z > \mid z_t\mid$, $w(z)$ would mimic the evolution obtained with the CPL $w_0$-$w_a$ parametrization. 
When $z_t=0$, $w_0+w_1 = w(0)$, and $w_1-w_0$ and $\Delta_z$ determine the slope of $w(z)$, i.e. it plays the same role as $w_a$. 
This corner of TDE parameter space would look different than $w(z) = w_0 + w_a z/(1+z)$ evolution at higher redshifts, but that is where the effect of dark energy on $H(z)$ is subdominant.

The TDE parametrization was introduced in Ref.~\cite{Keeley:2019esp} to explain the reconstructed $w(z)$ from a GP regression of the Planck CMB, Pantheon SN, and BOSS DR12 BAO.  
The TDE parametrization represented a viable solution to the $H_0$ tension, given the BAO datasets available at the time.  
This kind of late-time solution to the $H_0$ tension is no longer viable; the BOSS DR16 BAO dataset ruled it out as an explanation for the $H_0$ tension \cite{Keeley:2022ojz}. 
With the new DESI BAO dataset, it is worth redoing the fits with the TDE parametrization to see what insights the new data may provide.   
Generally speaking, the TDE parametrization is more flexible than $w(z) = w_0 + w_a z/(1+z)$. 
In particular, deviations from $\Lambda$CDM or a constant equation of state, $w$CDM, can happen at redshifts other than $z=0$, which is not the case in the CPL parametrization for which $dw(a)/da$ is a constant. 
More generally, the TDE parametrization can serve as a framework that maps onto the true model, whatever that may be. To elaborate, with a sufficiently flexible parametrization, then ``phenomenological features'' might be used to point the way to a physical model.  For instance, we are finding that the DE density has to increase until $z\sim0.5$ then decrease.  This sort of phenomenological feature might point the way to the true model, which would also have this phenomenological feature, and thus TDE would ``map'' onto the true model.

\section{The preference for evolving dark energy }\label{sec:results}

\begin{figure}
    \centering
    \includegraphics[width=\columnwidth]{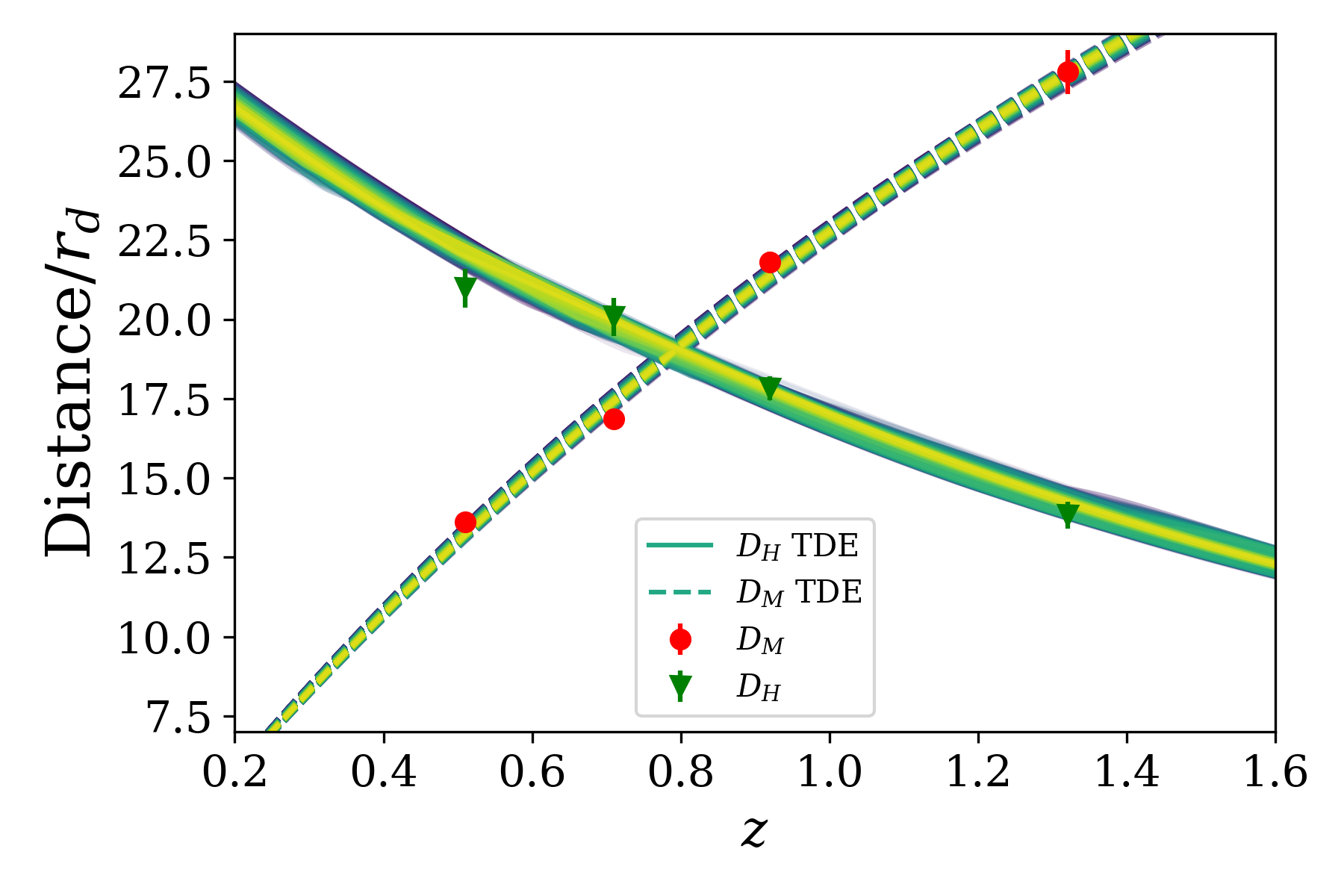}
    \caption{Posterior samples of $D_H(z)$ and $D_M(z)$ from the joint fit to the DESI BAO, 5-year DES SN, and Planck CMB datasets with the DESI data plotted on top of them.}
    \label{fig:TDE_dist}
\end{figure}

\begin{figure*}
    \centering
    \includegraphics[width=\columnwidth]{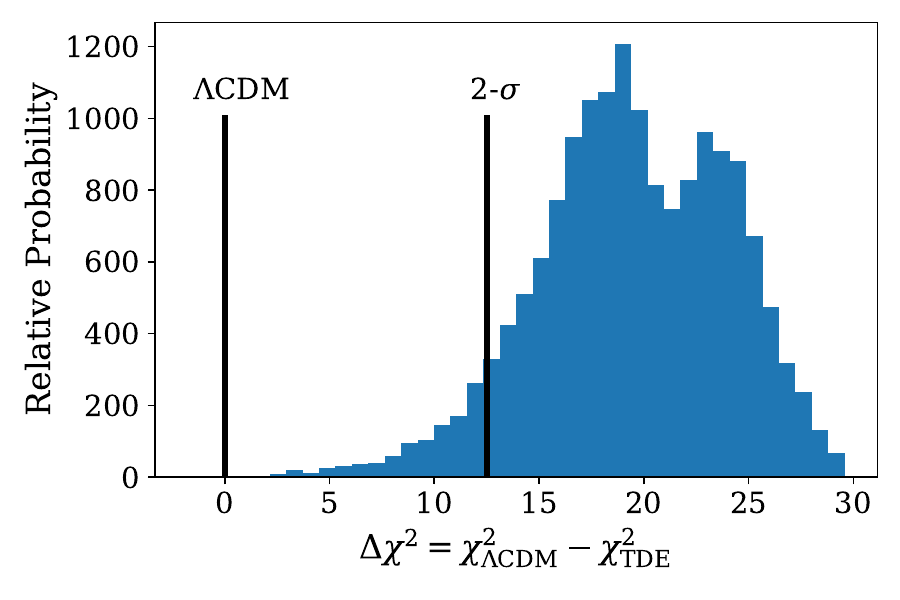}
    \includegraphics[width=\columnwidth]{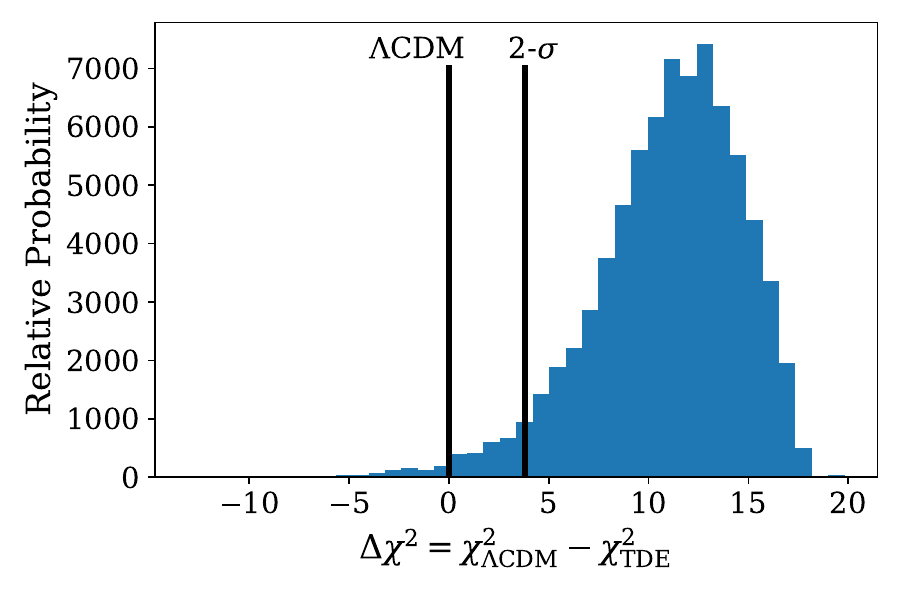}
    \caption{Distribution of the $\Delta \chi^2$ values from the posterior of the TDE fit to the DESI BAO, 5-year DES SN, and Planck CMB datasets, relative to the best-fit $\Lambda$CDM model to the same datasets.  $\Lambda$CDM sits in a very unlikely portion of the $\Delta \chi^2$ distribution, relative to the TDE parametrization.}
    \label{fig:TDE_dchi2}
\end{figure*}

Here, we present results for fitting the TDE parametrization to the joint CMB+BAO+SN datasets. We investigate the preference for evolving dark energy with both the DES and PP SN datasets. In Figure~\ref{fig:TDE_dist}, we show the posterior predictive distribution of $D_H(z)$ and $D_M(z)$ from TDE fits to the CMB+DESI+DES dataset, which illustrates the DESI constraints. The posterior predictive distribution is just the functions $D_H(z)$ and $D_M(z)$ that correspond to each individual set of parameters from the Markov Chain Monte Carlo fit to the data. Individual samples are color-coded by their $\chi^2$ with yellow being the best fit and blue being $\Delta \chi^2 = 15$, a 2-$\sigma$ deviation for our 8 parameter model (see Fig.~\ref{fig:TDE_dchi2}).

\begin{figure*}
    \centering
    \includegraphics[width=\columnwidth]{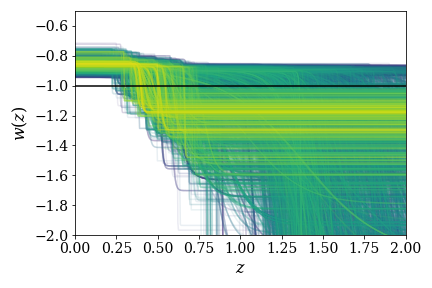}
    \includegraphics[width=\columnwidth]{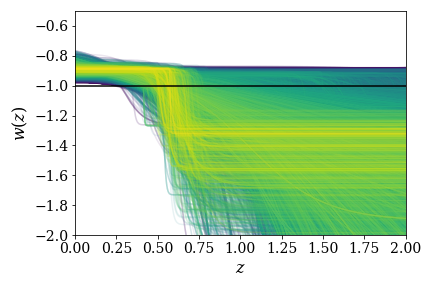}\\
    \includegraphics[width=\columnwidth]{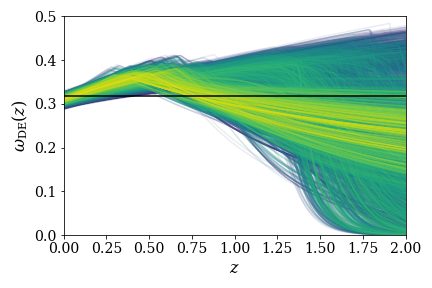}
    \includegraphics[width=\columnwidth]{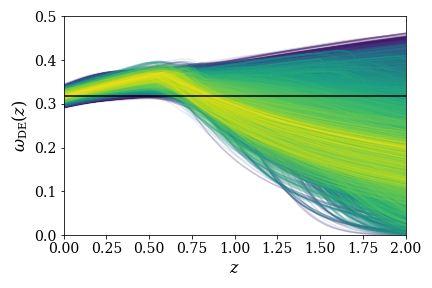}\\
    \includegraphics[width=\columnwidth]{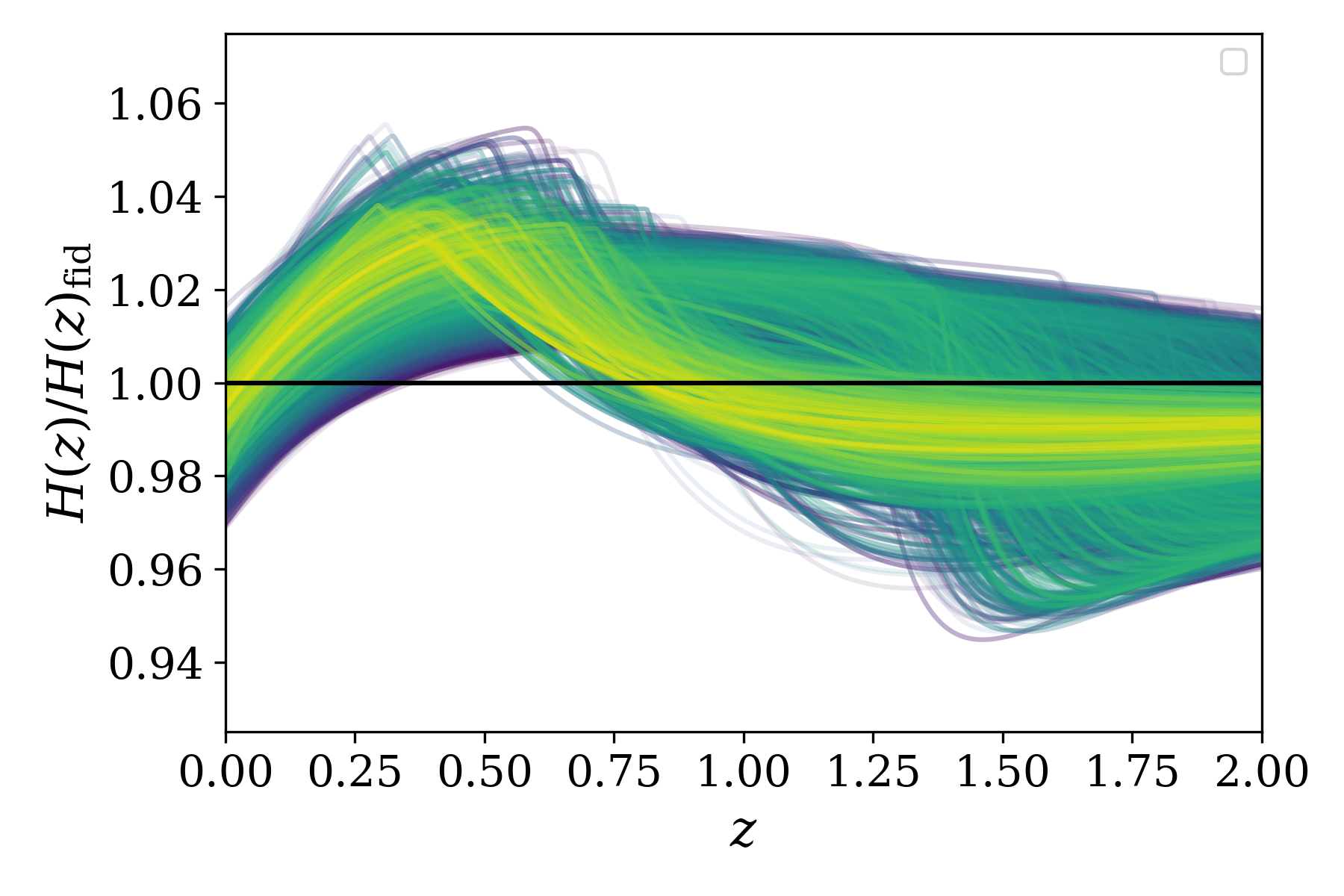}  
    \includegraphics[width=\columnwidth]{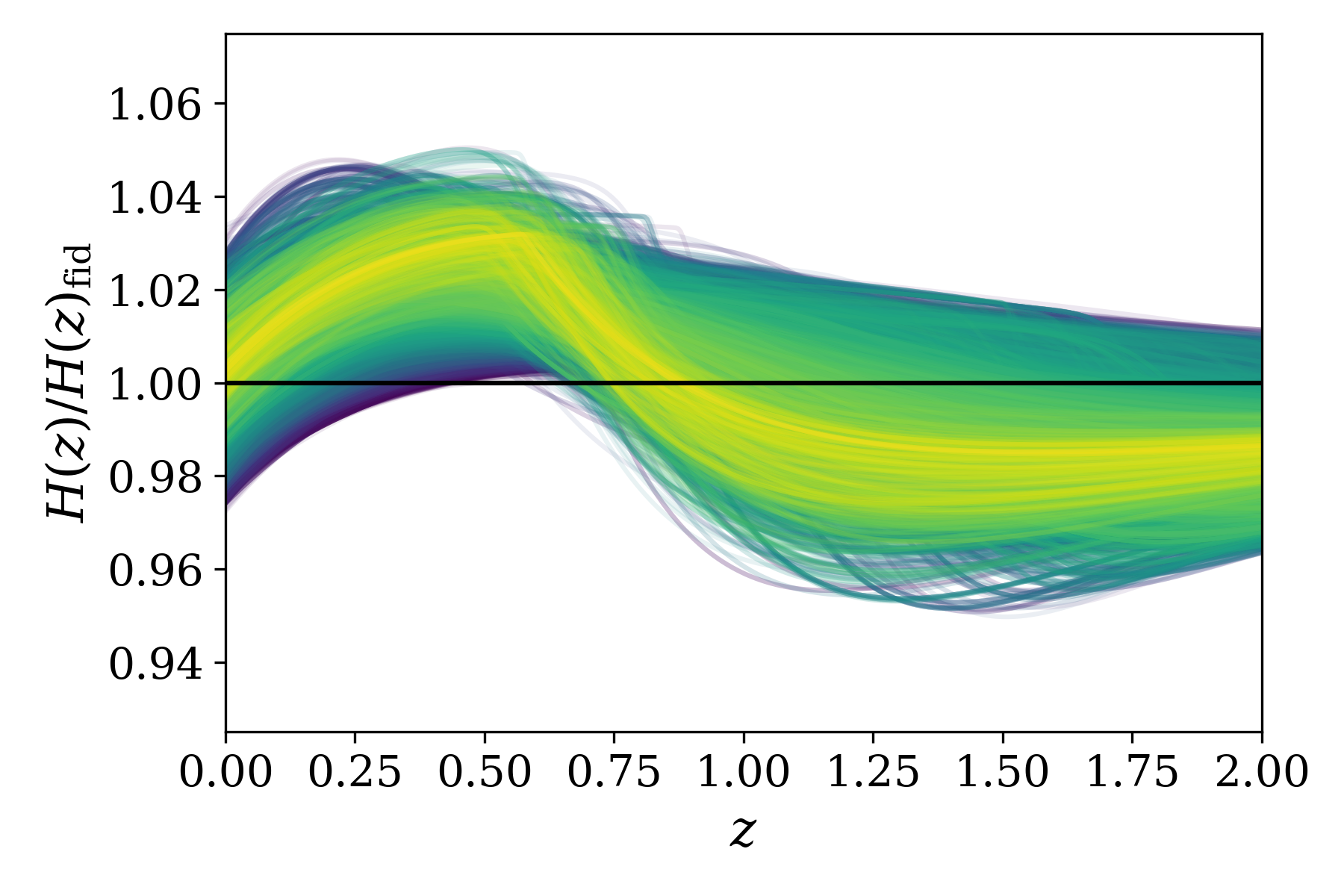}
    \caption{Posterior samples of $w(z)$ for the case with the joint fit to the Planck CMB, DESI BAO, and 5-year DES SN (left) and for the case with Planck CMB, DESI BAO, and PP SN (right) are shown. Each of the drawn functions is color-coded by their fit to the data, with yellow being the best fit, and blue being the 2-$\sigma$ deviation from the best fit. Each sample shows an evolution in $w(z)$. A black line at $w(z)=-1$ is shown to represent $\Lambda$CDM for comparison. The middle panels show the corresponding $\omega_{\rm DE}(z)$ posteriors. The bottom panels show the corresponding $H(z)$ divided by the Planck-$\Lambda$CDM model. These functions are more useful for building intuition about how the expansion history affects the growth history (see Sec.~\ref{sec:growth}).}
    \label{fig:TDE_wz_post}
\end{figure*}

In Figure~\ref{fig:TDE_wz_post}, we show the posterior predictive distribution of the dark energy equation of state $w(z)$ for the two cases of using the DES and then the PP SN datasets in combination with the CMB and DESI datasets.
The color coding is the same as in Figure~\ref{fig:TDE_dist} where yellow is the best fit and blue is a 2-$\sigma$ deviation as calculated in Fig.~\ref{fig:TDE_dchi2}. 
The CMB+DESI+DES datasets prefer equations of states that are quintessence-like at $z=0$ with $w(z=0) = -0.85 \pm 0.05$ ($w(z=0) = -0.90 \pm 0.05$ for CMB+DESI+PP).  For both SN datasets, the equation of state consistently shifts to a more phantom-like value at redshifts above $z>0.25$; however, the exact redshift at which this transition occurs, as well as the value to which it shifts, remains poorly constrained. Correspondingly, the dark energy density starts below the value from the best-fit Planck-$\Lambda$CDM at redshifts above the transition redshift, turns on at the transition redshift to values above the Planck-$\Lambda$CDM values, then evolves to smaller values as $w(z)$ is quintessence-like after the transition.

In Figure~\ref{fig:TDE_dchi2}, we show the distribution of the $\Delta \chi^2$ values from the posterior of the TDE fit to the joint DESI BAO and Planck CMB datasets, alongside both SN cases, the 5-year DES SN dataset on the left, and with the PP dataset on the right, relative to the best-fit $\Lambda$CDM model ($\Delta \chi^2 = \chi^2_{\rm TDE} - \chi^2_{ \Lambda \rm{CDM}}$).  
That the best-fit $\Lambda$CDM $\chi^2$ value falls in the tail of this distribution indicates that the TDE parametrization fits the joint datasets significantly better than the best-fit $\Lambda$CDM model. 
For the 5-year DES SN dataset, 99.99\% of the distribution of the $\chi^2$ values falls to the right of the best-fit $\Lambda$CDM model indicating at 3.8~$\sigma$ preference for TDE over $\Lambda$CDM. 
This deviates from a direct calculation of the square-root of the total $\Delta \chi^2 = 29.8$ of the best-fit $\Lambda$CDM model with respect to the best-fit TDE parametrization due to the non-Gaussian nature of the likelihood. 
A contribution to the $\Delta \chi^2$ of 17.8 is coming improvements in the SN likelihood, and a contribution of 12.0 is coming from improvements in the BAO likelihood, while the CMB preserves the same fit.  
For the PP dataset, the preference for TDE over $\Lambda$CDM is less significant at 2.4$\sigma$. 
Further details of the fits are available in Table~\ref{tab:chi2_table}.

\begin{figure}
    \centering
    \includegraphics[width=0.5\textwidth]{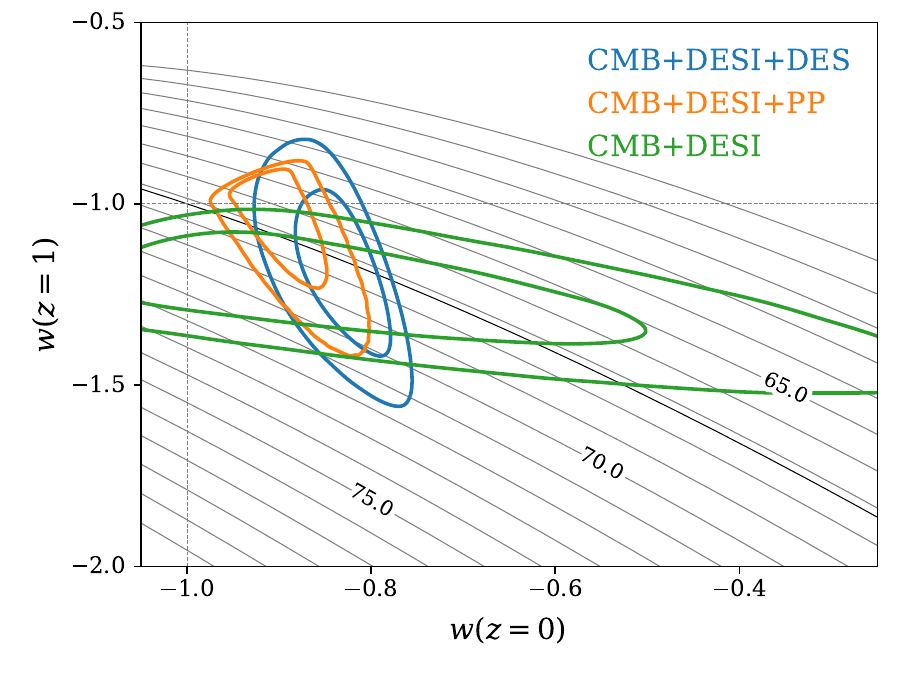}
    \caption{The posterior of $w(z=0)$ vs $w(z=1)$ for the TDE parametrization fit to the the DESI BAO, SN, and Planck CMB datasets (blue contours: DES SN, orange contours: PP SN) is shown, with 68\% and 95\% curves. The black line shows values of $w(z=0)$ and $w(z=1)$ (from a $w_0\, \&\,w_a$ model) that satisfy the constraint $H_0= 67.36\,\mathrm{km\,s^{-1}\,Mpc^{-1}}$, $\Omega_m h^2=0.14237$, and $100\theta_* = 1.04092$ (best-fit values from Planck). The grey curves satisfy $\Omega_m h^2=0.14237$, and $100\theta_* = 1.04092$ but $H_0$ spans $80\,\mathrm{km\,s^{-1}\,Mpc^{-1}}$ to $60\,\mathrm{km\,s^{-1}\,Mpc^{-1}}$.}
    \label{fig:mirage}
\end{figure}

Both SN datasets favor more dark energy than the $\Lambda$CDM model at redshifts between about 0.25 and 0.75, though the evidence for this is less significant in the PP dataset. This is consistent with previous works which found consistency between different SN datasets~\cite{2025JCAP...01..064M}  Recall that, if a cosmology is constrained by the CMB, then it needs to give the right angular size subtended by the sound horizon. Assuming no change to the sound horizon at last scattering, this is a tight constraint on the angular diameter distance to the last scattering surface $D_A(z_*)$. This can produce a ``mirage'' of a $w=-1$ dark energy while having a strongly evolving DE, subject to certain constraints \cite{Linder:2007ka}. The mirage requires that, in order to satisfy the distance to the surface of last scattering, for models where $w(z=0)>-1$, then $w(z)$ must be $<-1$ at higher redshifts, and vice versa.  The CMB constraint alone selects a family of $w(z)$ functions that have this feature. However, both of the SNe datasets show a preference for $w(z=0)>-1$ and so $w(z>1)<-1$ is preferred. In Fig.~\ref{fig:mirage}, we show the posterior of $w(z=0)$ and $w(z=1)$ for both the CMB+DESI+DES and CMB+DESI+PP joint datasets, along with a curve (in black) through that space that represents the values of $w(z=0)$ and $w(z=1)$ that satisfy the best-fit Planck parameters for the background expansion ($H_0= 67.36\,\mathrm{km\,s^{-1}\,Mpc^{-1}}$ $\Omega_m h^2=0.14237$, and $100\theta_* = 1.04092$). 

We also show the curves with fixed $H_0$ that keep $D_A(z_*)$ constant in the $w(z=0)$--$w(z=1)$ plane for the CPL parametrization to demonstrate the impact of the CMB and how the mirage appears. 
We have used the CPL parametrization because it uniquely specifies a curve in this plane for fixed $H_0$; the CMB-only constraints for the TDE parametrization follow the CPL curves.  
Further, when $H_0$ is close to $67.36\,\mathrm{km\,s^{-1}\,Mpc^{-1}}$, this calculation yields the curve of $w(z=0)$, $w(z=1)$ values that would most closely mimic $\Lambda$CDM without including high-quality measurements of low-redshift distances from SN or BAO. 
For values of $H_0$ that are larger than about $65\,\mathrm{km\,s^{-1}\,Mpc^{-1}}$, most of the curves lie in the region of $w(z=0)>-1$ and $w(z=1)<-1$, i.e., phantom crossing is generic.  
This plot---and a variation of it---is useful for assessing the origin of the preference for TDE or evolving dark energy, particularly in relation to specific datasets.  

The CMB alone allows for entire families of $w(z=0)$, $w(z=1)$ curves.  
The preference for evolving dark energy arises from a high-quality low-redshift measurement, for example, see Figure \ref{fig:No-DESI} for the combination of DES and CMB or PP and CMB. 
The SN datasets prefer $w(z=0)>-1$, the BAO dataset, calibrated by the CMB's sound horizon, prefers $70>H_0 > 65\,\mathrm{km\,s^{-1}\,Mpc^{-1}}$ and the CMB's angular diameter distance sets the evolution of $dw/da$. 
The constant $w(z)$ model still wants $H_0\sim 67.7$ km/s/Mpc \cite{DESI:2024mwx}. 
Thus the CMB constraint puts us along the black curve that intersects close to $(w(z=0),w(z=1))=(-1,-1)$. 
If by assumption, $w_a$ is forced to be zero (i.e the constant $w(z)$ model), then the CMB alone would want $w_0\sim-1$. 
This seems to be a coincidence, and the fact that constant $w(z)$ models prefer $w=-1$ does not by itself illuminate the question of whether the CMB+DESI+SN preference for evolving dark energy is robust. 
It is worth noting in this regard that the DESI and SN contours have different slopes and neither follows the CMB curves faithfully, demonstrating the fact that they are adding independent information about dark energy changing with time.

In our analyses, we see that significant improvements in the fit to the SN data occurs at redshifts $z>0.2$, as shown in Fig.~\ref{fig:binned_SN}. 
It is also at $z>0.2$ where the TDE parametrization improves the fit to the DESI BAO dataset, as shown in Fig.~\ref{fig:BAO_diag}.  
The DES SN and the DESI BAO datasets, and to a less significant extent the PP SN dataset, shift the model's distances in the same direction (preferred distances are smaller) at $z>0.2$. 
In other words, the same change in the expansion history (between TDE and $\Lambda$CDM) at redshifts $z>0.2$ is simultaneously improving both the SN and BAO datasets.
Although this preference for evolving dark energy could be due to systematic effects given that the level of significance in the PP and DESI data sets (when combined with the CMB) is not high, it must be borne in mind that this is also what one would expect from new physics. 

\begin{figure}
    \centering
    \includegraphics[width=\columnwidth]{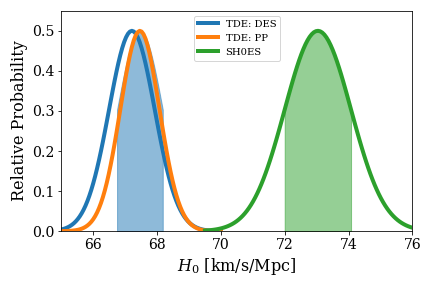}
    \caption{The posterior distribution for $H_0$ when fitting the TDE parametrization to the joint CMB+BAO+DES dataset (blue) and CMB+BAO+PP dataset (orange) is shown. Models that add new physics at low-redshift cannot solve the $H_0$ tension.}
    \label{fig:TDE_H0}
\end{figure}

In Figure~\ref{fig:TDE_H0}, we plot the posterior of the derived parameter $H_0$. Here, $H_0 = (67.12 \pm 0.72)\,\mathrm{km\,s^{-1}\,Mpc^{-1}}$. Despite the additional flexibility of the TDE parametrization, the expansion histories that fit the DESI BAO data well do not shift the $H_0$ posterior towards the SH0ES measurement of $H_0$~\cite{Riess:2021jrx}, and therefore do not solve the $H_0$ tension.

\section{Assessing the preference with the growth history}\label{sec:growth}

The standard equation for the growth of matter perturbations ($\delta_m$) in General Relativity \cite{Dodelson:2020bqr} is given by, 
\begin{equation} \label{eq:deltam}
a^2 \delta_m''(a) + {d\ln(a^3H(a)) \over d\ln(a)} a \delta_m'(a) - \frac{3}{2} \Omega_m(a) \delta_m(a) = 0\,,
\end{equation}
where $\Omega_m(a) \equiv \omega_m (1+z)^3/ h(z)^2$. This assumes that only the matter component clusters on small scales so that the gravitational potential $\phi \propto \delta_m/a$, and that there is no coupling between dark matter and dark energy. The equation is derived by assuming the energy-momentum conservation of the dark matter fluid (with the usual perturbed metric) and the Poisson equation on small scales.  Since the expansion rate directly impacts the growth of perturbations in this equation, an independent inference of the growth rate allows for a consistency check with the distance measurements in the context of GR. 

\newcommand{\dbm}{\delta_{\rm bm}}
\newcommand{\Hbm}{H_{\rm bm}}
Given that the deviations from a baseline Planck-$\Lambda$CDM model (which we designate as ``bm'') are restricted by data to be less than approximately 2.5\% at $z<2$, the deviations in the growth function, in the context of the above equation, will also be restricted. To explore the dependence of the deviation away from the baseline Planck model, we can write the matter perturbation as $\delta_m(a) = \dbm(a) (1+u(a))$. We can then write the approximate equation assuming $u(a) \ll 1$ as: 

\begin{eqnarray}
{d^2 u \over d \ln(a)^2} &+& \left(2 + 2\Omega_m(a)^\gamma - 3 \Omega_m(a)/2\right) {d u \over d \ln(a)} \nonumber \\
&+&  \Omega_m(a)^\gamma \xi(a) - \frac{3}{2} \delta \Omega_m(a) = 0\,,\label{ugrowth}
\end{eqnarray}
where 
\begin{equation}
\xi(a) \equiv \frac{d\ln(H(a)/\Hbm(a))}{d\ln(a)} \approx \frac{d(\delta H(a)/\Hbm(a))}{d\ln(a)} \,, 
\end{equation}
and $\xi(a) \ll 1$. We have used the approximation $d\ln(\dbm(a))/d\ln(a) = \Omega_m(a)^\gamma$ with $\gamma=0.55$ for the baseline $\Lambda$CDM model. The last term can be written as $\delta \Omega_m(a) \approx \Omega_m(a) (\delta w_m/w_m -2 \delta H(a)/\Hbm(a))$ for small variations in GR but it can also accommodate deviations from GR that change the effective Newton's constant in the Poisson equation for linear perturbations. For example, if the effective gravitational constant is $G_{\rm eff}(a)$ then it will contribute the $\delta \Omega_m(a)$ term as $(G_{\rm eff}(a)/G_N-1)\Omega_m(a)$. This means that the redshift dependence of the change in the growth rate due to $G_{\rm eff}(a) \neq G_N$ could be distinct from that due to $\delta H(a)$. 

Eq.~\eqref{ugrowth} can be integrated to give
\begin{widetext}
\begin{equation}
\frac{du(a)}{d\ln(a)}  = - \frac{\delta H(a)}{H(a)} \Omega_m(a)^\gamma + 
\int_{a_i}^a da'  \left( \frac{\delta H(a')}{H(a')} \frac{d (\Omega_m(a')^\gamma G(a',a))}{da'}  + \frac{3}{2} \frac{\delta \Omega_m(a') G(a',a)}{2 a'}  \right) \,,  \label{firstintegral}
\end{equation}
\end{widetext}
where we have assumed that $du/da = 0$ at $a=a_i$ and to a good approximation (for the purposes of doing the integral numerically) we can use $G(a',a) = (a'/a)^\beta$ with $\beta = 2.6$. More generally, $G(a',a) = \exp(\int_{a}^{a'} da' \beta(a')/a')$ where $\beta(a) = \left(2 + 2\Omega_m(a)^\gamma - 3 \Omega_m(a)/2\right)$ and it lies in the range $[2.5,2.616]$ for $\Omega_m$ in the range of 0.2 to 1. 

The equation above simplifies further because the contribution of the integral is small, and so it is a good approximation to simply set $du/d\ln(a) = - \Omega_m(a)^\gamma\,  \delta H(a)/H(a)$. 
This shows that changes in peculiar velocities (between two models) are directly sourced by the change in the expansion rate. Hence, measurements of redshift space distortions will provide a direct handle on the deviations of the expansion rate from a baseline $\Lambda$CDM model.  

Eq.~\eqref{firstintegral} can be integrated again to give
\begin{eqnarray}
u(a) = \int_{a_i}^a da' && \left( \frac{\delta H(a')}{H(a')} \frac{d (\Omega_m(a')^\gamma F(a',a) )}{da'} \right. + 
\nonumber\\ 
&& \left.  \frac{3}{2} \frac{\delta \Omega_m(a') F(a',a)}{a'} \right) \,,
\end{eqnarray}
where we $u(a_i) = 0$ and $dF(a',a)/d\ln(a')=G(a',a)$. 
With the approximation of treating $\beta$ as constant, we obtain $F(a',a) = (1-(a'/a)^\beta)/\beta$.

Putting these approximations together, we can find the changes to $f(a)\sigma_8(a)$ using $\delta'(a) = \dbm'(a) ( 1+ u(a) + \Omega_m(a)^{-\gamma}du(a)/d\ln(a))$. 
This allows us to write the fractional change in $f(a)\sigma_8(a)$ as $u(a)-u(1) + \Omega_m(a)^{-\gamma}du(a)/d\ln(a)$ for small changes. 
With the further approximation $\Omega_m(a)^{-\gamma} du(a)/d\ln(a) = -\delta H(a)/H(a)$, which comes from neglecting the second term on the RHS of Eq.~\ref{firstintegral}, we can write 
\begin{equation}\label{eq:delta-fsigma8}
\frac{\delta(f \sigma_8(a))}{f \sigma_8(a)} = -\frac{\delta H(a)}{H(a)} + \int_{a}^{1} da'  \frac{\delta H(a')}{H(a')} \frac{\Omega_m(a')^\gamma}{a'}\, .
\end{equation}
The $\delta H(a)/H(a)$ term drives the specific shape of the deviation $\delta(f \sigma_8(a))$ but the integral part of the right hand side of Eq.~\eqref{eq:delta-fsigma8} is still important in that it gives $0.5\%-1\%$ level corrections to $\delta(f \sigma_8(a))$. 
The full approximation in Eq.~\ref{eq:delta-fsigma8} when compared to solving Eq.~\ref{eq:deltam} is good to about 3\%. 

The fits to the distances shows a preference for more dark energy than predicted by the $\Lambda$CDM model in the sense of the equation of state $w(z) > -1$ for $z \lesssim 0.5$. 
This means that the inferred expansion rate is higher than $\Lambda$CDM prediction at $z \lesssim 0.5$. 
From the equation above, we see that (assuming GR) the prediction is that $f\sigma_8(z)$ is lower at $z \lesssim 0.5$.  

\begin{figure}
    \centering
    \includegraphics[width=\columnwidth]{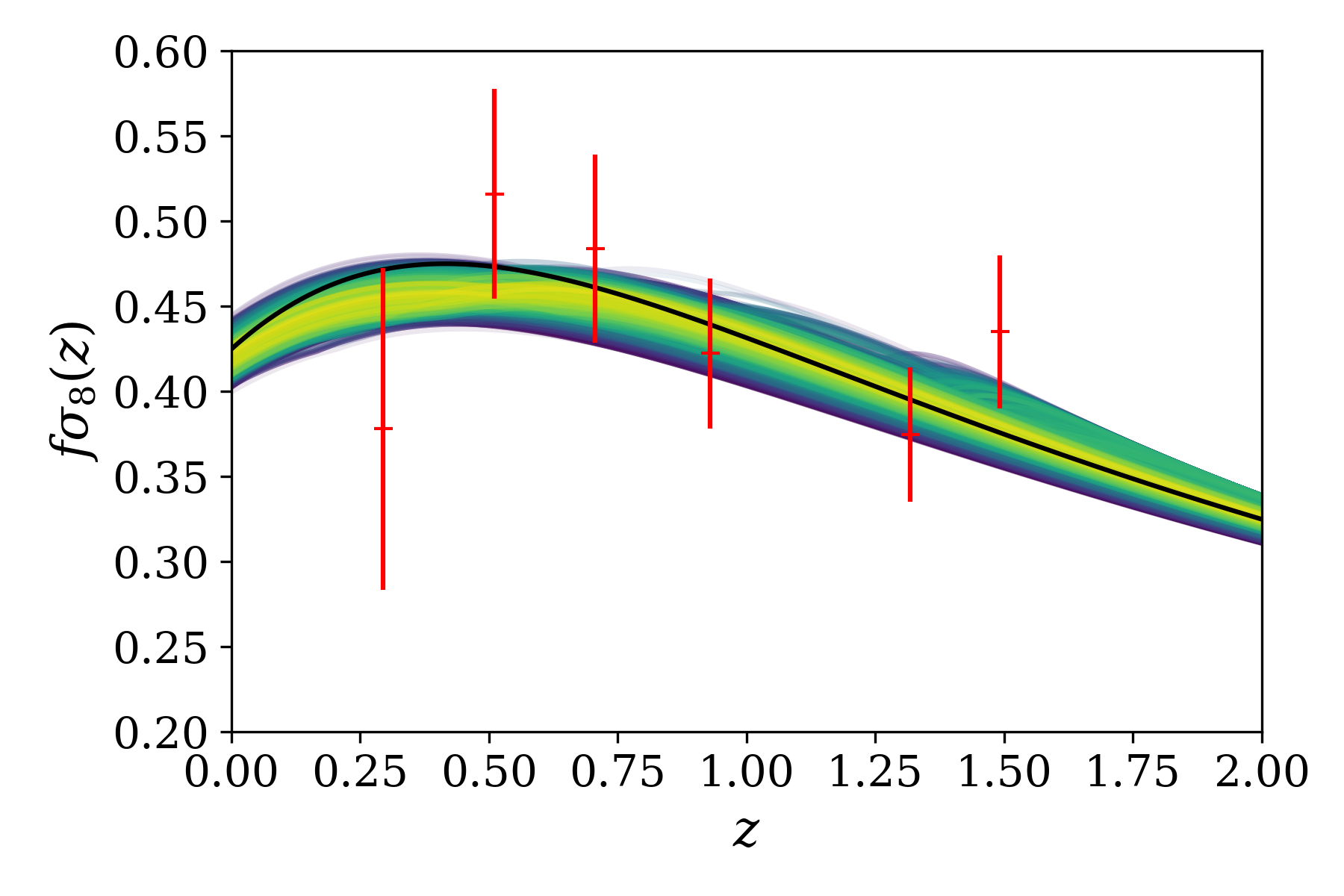}
    \includegraphics[width=\columnwidth]{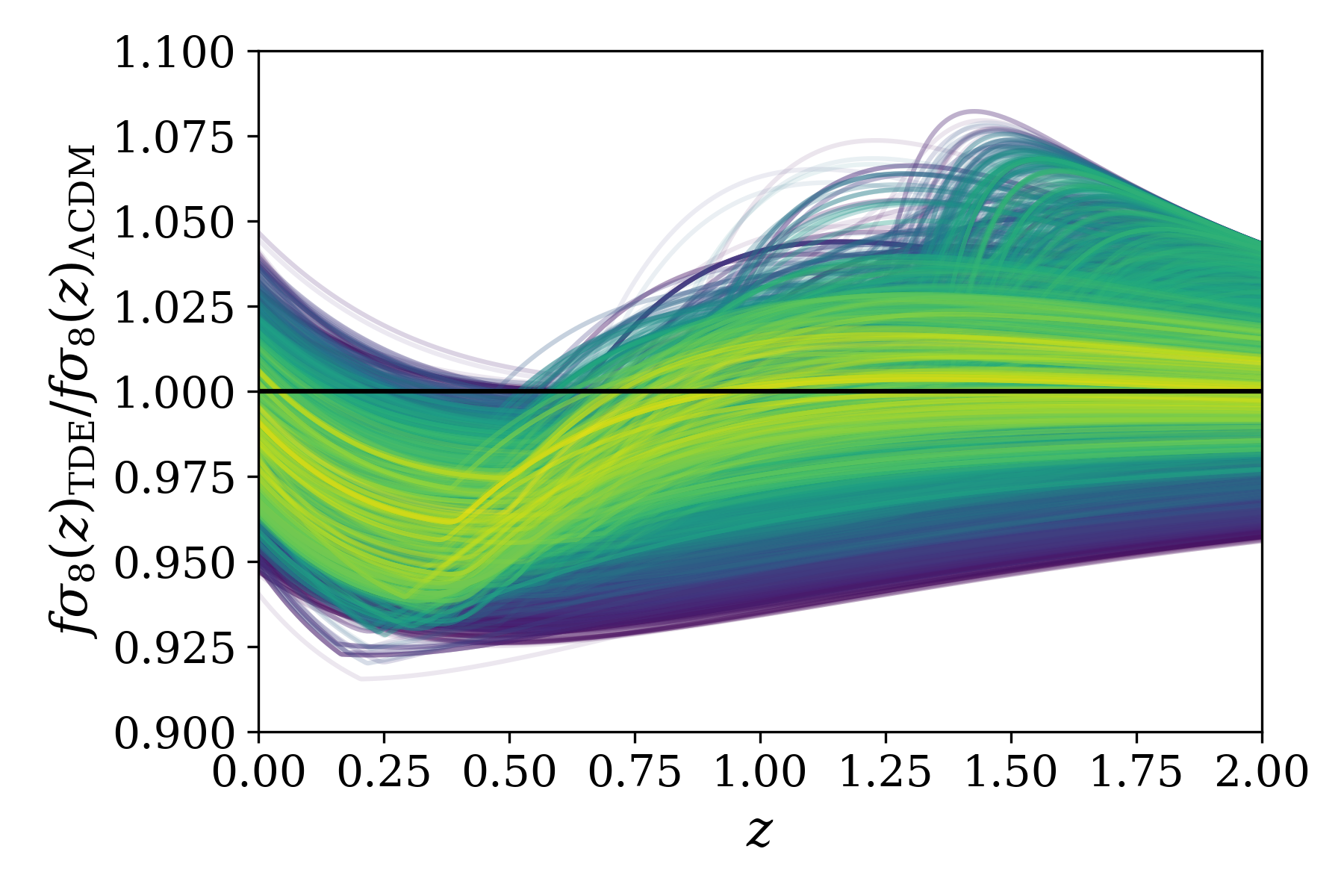}
    \caption{Top panel: Posterior samples of $f\sigma_8(z)$ from fitting the TDE parametrization to the joint CMB+BAO+DES dataset. The red points represent the DESI RSD data~\cite{DESI:2024jis}. Bottom panel: Same but for the ratio of $f\sigma_8(z)$ between TDE and $\Lambda$CDM. The DESI whitepaper~\cite{DESI:2016fyo} forecasts $\sim 3\%$ errors on $f\sigma_8$ so this predicted flattening in $f\sigma_8(z)$ may be detectable.}
    \label{fig:fs8}
\end{figure}

In Figure~\ref{fig:fs8}, we show the predicted distribution of $f\sigma_8(z)$ values drawn from the TDE posterior fit to CMB+BAO+SN datasets. 
Compared to the Planck-$\Lambda$CDM curve in black, at redshifts above $z>1$, the best-fit TDE parameters predict the same values for $f\sigma_8(z)$, but at $z<1$, the curve has a different shape.  
Between redshifts $0.25<z<0.75$ the dark energy density (and thus $H(z)$) is higher in the TDE than in the Planck-$\Lambda$CDM model. 
This decreases the growth rate of structure at those redshifts as predicted by Eq.~\eqref{eq:delta-fsigma8}.  
By redshift $z=0$, the dark energy density (and hence the expansion rate) of the TDE parametrization is smaller than in the Planck-$\Lambda$CDM model, causing the growth in the TDE parametrization to increase relatively.  
On the whole, this makes $f\sigma_8(z)$ flatter with respect to the best-fit Planck model. 
The DESI whitepaper (Table 2.3 of~\cite{DESI:2016fyo}) forecasts $\sim 3\%$ errors on $f\sigma_8(z)$, over multiple redshifts bins, in the redshift range where TDE diverges from $\Lambda$CDM. 
Therefore, importantly, this flattening of the $f\sigma_8$ curve may likely be detectable.

\begin{figure}
    \centering
    \includegraphics[width=\columnwidth]{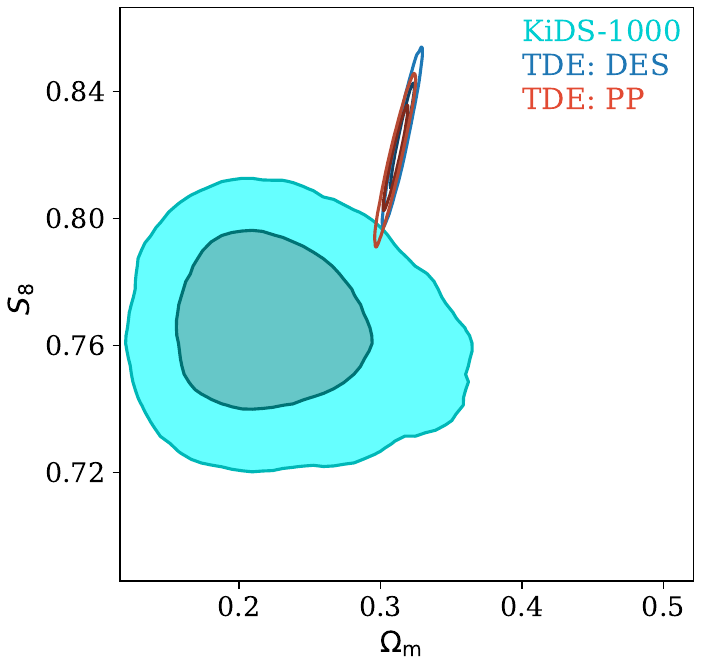}
    \caption{Posterior of $S_8-\Omega_{\rm m}$ from a TDE fit to the CMB+DESI+DES (blue) and CMB+DESI+PP (orange) datasets alongside the KiDS-1000 2PCFs constraint~\cite{KiDS:2020suj}. Note that the KiDS-1000 result are results for the $\Lambda$CDM model.}
    \label{fig:S8Om}
\end{figure}

In Figure~\ref{fig:S8Om}, we show the posterior of the parameters $S_8$ and $\Omega_{\rm m}$ from a TDE fit to the CMB+BAO+SN dataset. 
Individually, the constraint on these parameters are $S_8 = 0.826 \pm 0.011$ and $\Omega_{\rm m} = 0.3155 \pm 0.0055$. 
$S_8 = \sigma_8(\Omega_{\rm m}/0.3)^{0.5}$ is the combination of growth ($\sigma_8$) and expansion ($\Omega_{\rm m}$) that is best constrained by weak lensing datasets~\cite{Hildebrandt:2016iqg,KiDS:2020suj,DES:2021wwk}. 
The additional flexibility of the TDE parametrization, with respect to the Planck-$\Lambda$CDM model, shifts $S_8$ to lower values ($\Delta S_8 \sim 0.006$) while the errors are comparable. 
Ultimately, the expansion history that fits the DESI data (along with the CMB and SN) does not modify the growth history enough to solve the $S_8$ tension.

Other works have found that Planck, weak lensing, galaxy clustering, and cosmic velocity datasets preferred a suppressed growth of structure as parametrized by a larger growth index $\gamma$~\cite{Nguyen:2023fip}. This $\gamma$ parameter affects the growth of structure differently than the effects of evolving dark energy, like we calculate for the TDE model. Increasing $\gamma$ will only suppress the growth of structure, whereas evolving dark energy can have more complicated effects on $f\sigma_8(z)$.

\section{Discussion \& Conclusions}

Using a general parametrization of the dark energy equation of state (TDE), we find that an evolving dark energy model is preferred over the best-fit $\Lambda$CDM model at 3.8-$\sigma$ when using the DES SN dataset, and at 2.4-$\sigma$ when using the PP SN datasset, as shown in Figure~\ref{fig:TDE_dchi2}. 
We also calculate the Bayesian evidence ($Z$) and find that $\Delta \log Z = \log Z_{\rm TDE} - \log Z_{\rm \Lambda CDM} = 5.43$, when using DES, and $\Delta \log Z = 1.65$, when using PP.  
The preference for $w(z) \neq -1$ arises from the interplay of the different data sets; SN data sets prefer $w(z=0)>-1$, BAO and CMB sound horizon prefer $H_0 = 67$ km/s/Mpc and the CMB data determine how $w(z)$ evolves to match the angular diameter distance to the last scattering surface.

The preference we find for the TDE parametrization relative to $\Lambda$CDM comes from a $\Delta \chi^2 = 12.0$ improvement from the DESI BAO dataset and a $\Delta \chi^2 = 17.8$ improvement from the 5-year DES SN dataset.  It is important to emphasize that this preference arises from both the SN and BAO datasets. Indeed, the same modification to the expansion history at the same redshifts $0.25<z<0.75$ improves the fits to both the SN and BAO datasets. Therefore, if the preference were due to systematic uncertainties,  systematic effects would need to be present in both datasets.

The Planck dataset does not play a significant role in determining whether the TDE $w(z)$ or the $\Lambda$CDM model is preferred. However, it is essential for precisely constraining $\Omega_m$, which in turn allows deviations in the DESI BAO dataset from the Planck-$\Lambda$CDM model predictions to be interpreted as evidence of evolving dark energy rather than other modifications to the low-redshift expansion history. That is, the Planck CMB data breaks degeneracies between $\Omega_m$ and $w(z)$.
However, the discovery of the existence of systematic errors in either or both the DESI and 5-year DES SN datasets would reduce or eliminate the evidence for the TDE parametrization---or any evolving dark energy model---to fit the combined cosmological datasets better than $\Lambda$CDM~\cite{Keeley:2022iba,Efstathiou:2024xcq}. 

The TDE fit to these datasets finds $H_0 = (67.12 \pm 0.72)\,\mathrm{km\,s^{-1}\,Mpc^{-1}}$ and $S_8 = 0.826 \pm 0.011$, consistent with the Planck-$\Lambda$CDM inferred local expansion rate and clustering amplitude, and thus TDE does not solve the $H_0$ or $S_8$ tensions.
The freedom in the TDE parametrization of the dark energy equation of state is not sufficient to solve the $H_0$ tension, however there is a preference for the TDE parametrization over the $\Lambda$CDM model. Specifically, since TDE does not change the sound horizon, it cannot solve the $H_0$ tension with present BAO datasets.
In order to solve the $H_0$ and the $S_8$ tensions, given previous BAO datasets, we would need sharp transition in the equation of state of dark energy at redshift ($z_T$) of about unity \cite{Keeley:2019esp}. At the time, the BAO were flexible enough to allow for this to work for the $H_0$ tension without needing to change the sound horizon. eBOSS, with new intermediate QSO BAO data points that measured the expansion history where the transition was happening, tighter Ly$\alpha$ BAO measurements that shifted towards $\Lambda$CDM predictions, changed the story and no longer permitted this kind of solution. 
In order to explain the current SN and DESI BAO datasets, TDE wants a transition to occur at much lower redshifts.

An independent probe of the expansion rate, in the context of dark energy models based in General Relativity, is the rate of growth of linear perturbations. 
We have shown that for small variations in the expansion rate around the $\Lambda$CDM model, the allowed changes to the $f\sigma_8(z)$ curve track that of the expansion rate (see Eq.~\ref{eq:delta-fsigma8}). 
A key finding in our results is that a flattening of the $f\sigma_8(z)$ curve, compared to the $\Lambda$CDM prediction, in upcoming DESI RSD data would provide independent support for models of evolving dark energy that are preferred by the DESI and SNe data. 
The change in the growth rate leading to a flattening of the $f\sigma_8(z)$ curve is a signature of evolving dark energy density, contingent on General Relativity.

Another takeaway from our analysis is that the preference for evolving dark energy in the DESI and SN data sets (more prominently in DES) is not a result specific to the CPL parametrization. 
The fits using the TDE parametrization to the joint CMB+BAO+SN dataset shows that $w(z)$ pulls away from the $\Lambda$CDM expectation at redshifts where the data constrain the expansion history strongly.  
At $z < 0.5$, the dark energy equation of state $w(z) > -1$ is preferred, and going to higher redshifts there is a  trend for $w(z)$ to be close to or below $-1$. 
We note though that the constraints are not strong enough to exclude $w(z) > -1$ at $z=1$, so there is support for models of dark energy based on canonical scalar fields.
Our results provide concrete hints for ways in which future data on the expansion and growth rates could definitively reveal that dark energy evolves with time.

\begin{acknowledgments}
KNA and MK are partially supported by the U.S.
National Science Foundation (NSF) Theoretical Physics Program PHY-2210283. 
A.S. would like to acknowledge the support by National Research Foundation of Korea 2021M3F7A1082056, and the support of the Korea Institute for Advanced Study (KIAS) grant funded by the government of Korea.
\end{acknowledgments}

\appendix
\section{Diagnosing DESI BAO data}
In this section, we perform some tests that diagnose where the preference for an evolving dark energy model arises.

\begin{figure*}
\centering
\includegraphics[width=\columnwidth]{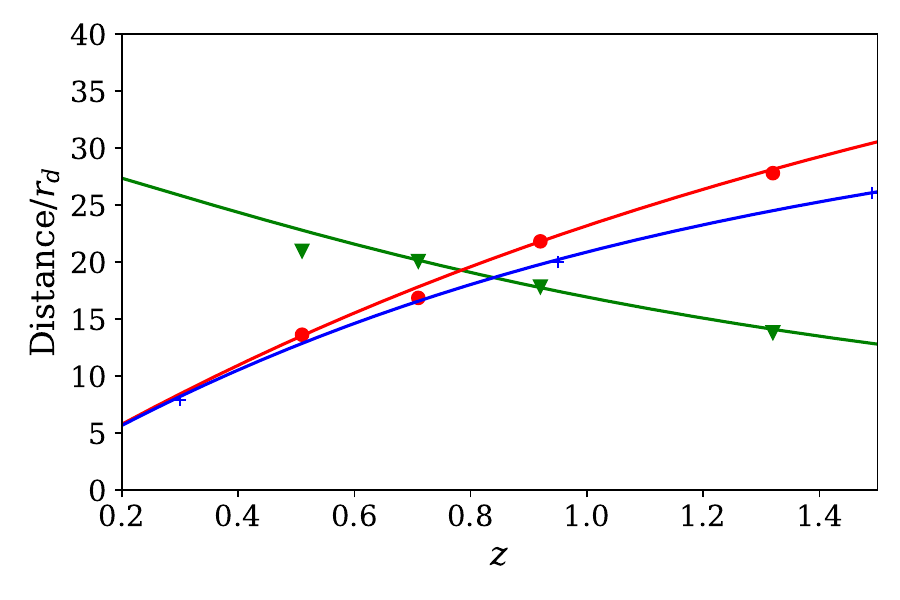}
\includegraphics[width=\columnwidth]{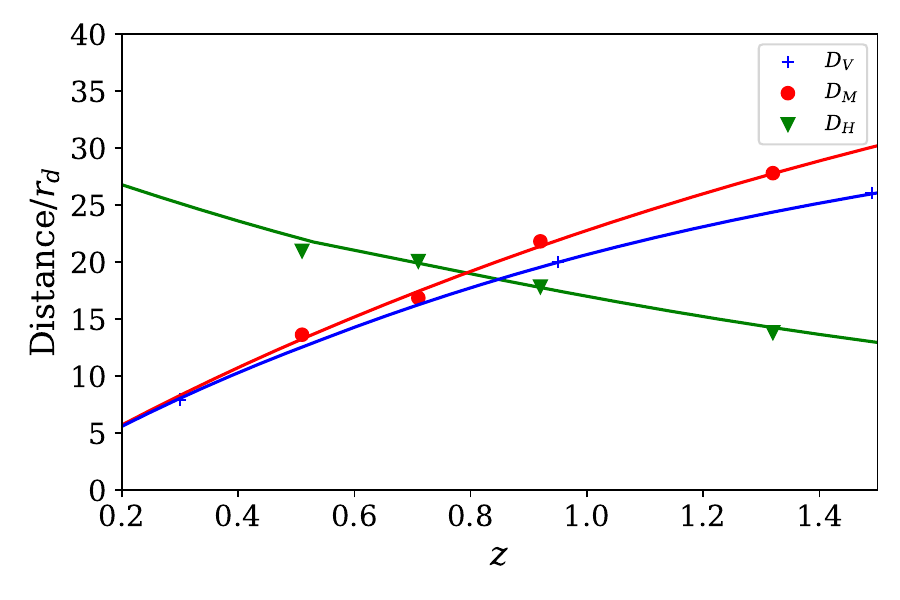}\\
\includegraphics[width=\columnwidth]{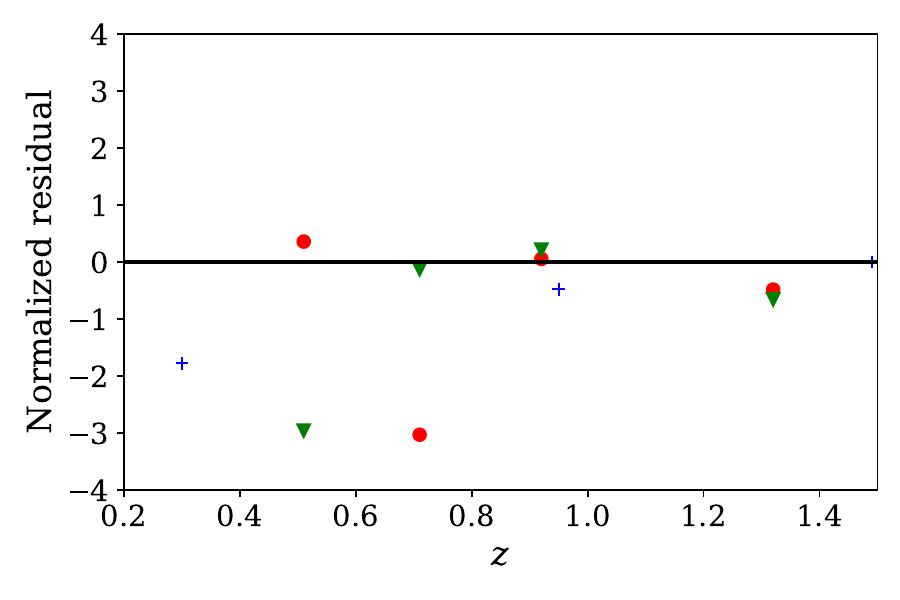}
\includegraphics[width=\columnwidth]{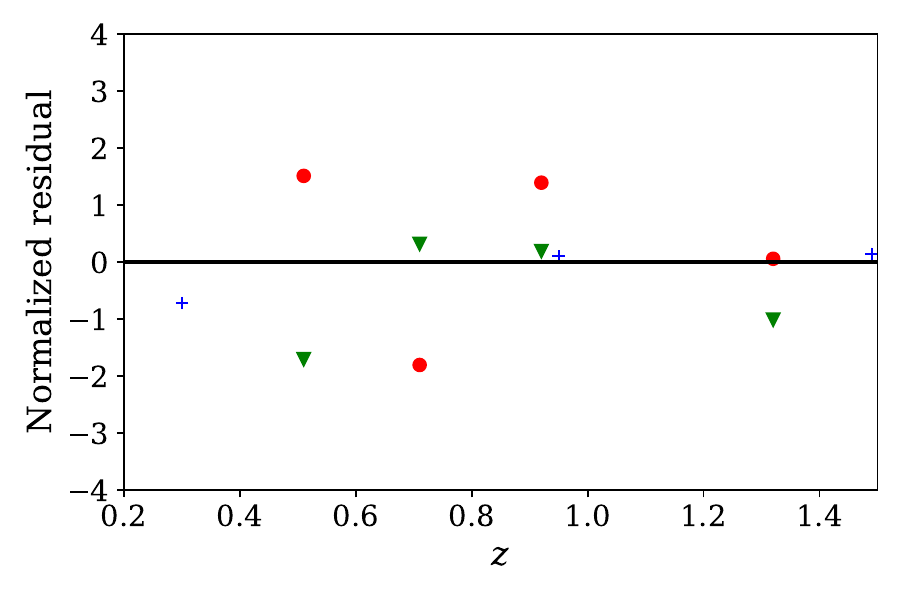}
\caption{Plots of the DESI data compared to the best-fit $\Lambda$CDM (left) and best-fit TDE (right). No one data point is driving the preference for TDE.}
\label{fig:BAO_diag}
\end{figure*}

Some works~\cite{Dinda:2024kjf} have argued that the DESI preference for evolving dark energy arises from a single data point (the $D_H$ point at $z=0.51$) in the DESI BAO dataset. In Figs.~\ref{fig:BAO_diag}, we compare the DESI data to the best-fit $\Lambda$CDM and TDE models to diagnose which data points are driving the preference. Contrary to the claims of ~\cite{Dinda:2024kjf}, multiple data points contribute to the preference and on the whole the residuals of the data points are much closer to a normal distribution with the TDE parametrization than the $\Lambda$CDM model.  The bottom-left-most points in the bottom-left panel of Fig.~\ref{fig:BAO_diag} are not outliers, especially compared to the best-fit TDE parametrization and excluding those points (which would be improper) would make the DESI data points very overfit. In other words, the datapoints would not be scattered away from the best-fit model to the extent as would be expected from their error bars; the number of ``$\sigma$''s would be too low compared to the number of datapoints.

\begin{figure*}
    \centering
    \includegraphics[width=\columnwidth]{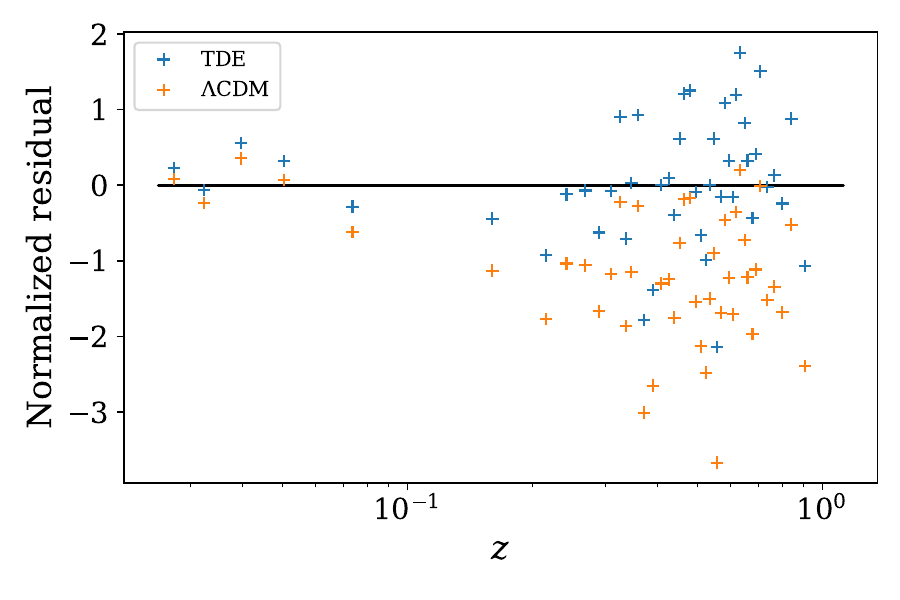}
    \includegraphics[width=\columnwidth]{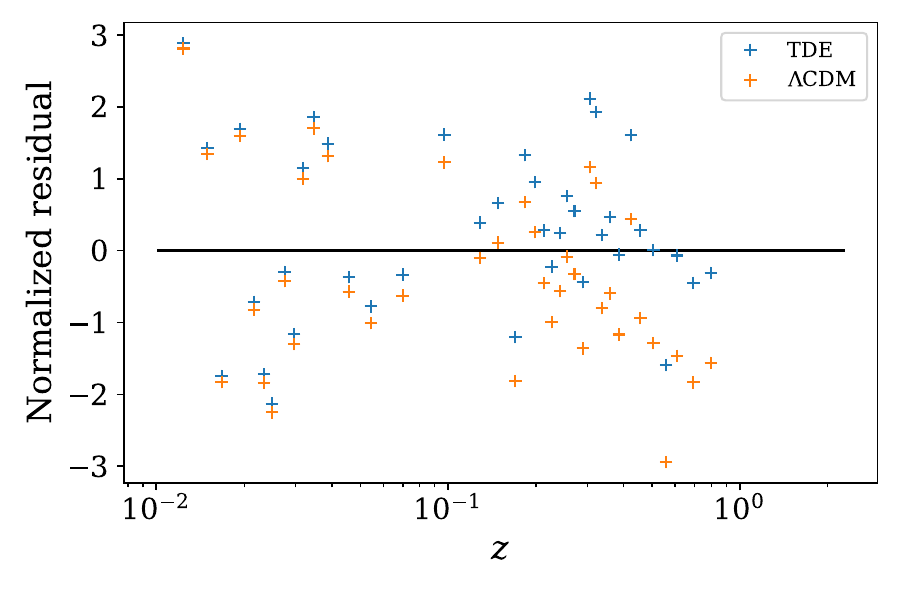}
    \caption{Binned normalized residuals for both the best-fit TDE (blue) and $\Lambda$CDM (orange) models for both the DES (left) and PP (right) datasets. It can be seen that the TDE parametrization shifts the predicted $m_b$(z)s to smaller values, especially at higher redshifts ($z\sim0.5-1$), and makes the normalized residuals more consistent with a normal distribution. }
    \label{fig:binned_SN}
\end{figure*}

In Fig.~\ref{fig:binned_SN}, we plot the normalized residuals for the best-fit TDE and best-fit $\Lambda$CDM models for the SN data from a joint CMB+DESI+SN fit. We show both the DES and PP cases.  The $\Lambda$CDM model yields SN residuals that are shifted and skewed with respect to a normal distribution and this is resolved by the additional flexibility of the TDE parametrization.  The residuals of the TDE parametrization are more consistent with a normal distribution.  The TDE parametrization works on the SN datasets by shifting the predictions for $D_L(z)$ (and hence $\mu(z)$ and $m_b(z)$) to lower values, especially at higher redshifts ($z\sim0.5-1$). This shift at higher redshifts is the effect of the evolution of the dark energy density.

\begin{figure*}
    \centering
    \includegraphics[width=\columnwidth]{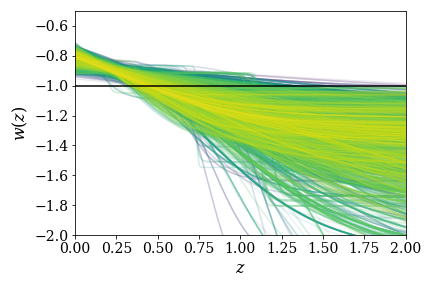}
    \includegraphics[width=\columnwidth]{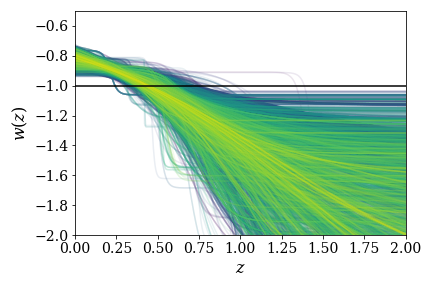}
    \caption{Like Fig.~\ref{fig:TDE_wz_post} but for the case where the  $D_H$ data point at $z=0.51$ is excluded in the left panel. In the right panel, the full DESI dataset is included and, in addition, the SH0ES constraint on $H_0$ is included.}
    \label{fig:noz05}
\end{figure*}

To explicitly test this claim, that the (the $D_H$ point at $z=0.51$) is not the sole data point driving the preference for evolving dark energy, we rerun the analysis by excluding this data point in the likelihood.  
The results are shown in the left panel of Fig. ~\ref{fig:noz05}.  
The specific shape of the posterior predictive distribution is different but there is still a preference for $w(z) > -1$ 
at $z<0.5$ and phantom-like behavior at $z>0.5$.  Interestingly, the absence of the $z=0.51$ data point reduces the preference for a sharp transition, and returns $w(z)$ curves similar to those of $w_0, w_a$ parametrizations.  As expected, the significance is reduced from 3.8 to 3.1 $\sigma$. 

We also test the case where the SH0ES $H_0$ constraint~\cite{Rubin:2023ovl} is also included in the joint likelihood. Note that this is not a consistent way to include the $H_0$ constraint~\cite{Efstathiou:2021ocp}, but it suffices for our limited purpose here to investigate what impact this would have. The results of this case is shown in the right panel of Fig.~\ref{fig:noz05}. The preference for an evolving dark energy is still present even though the specific shape of the evolution is different.  The best-fit value of $H_0$ shifts to $H_0 = (68.2\pm 0.5)\,\mathrm{km\,s^{-1}\,Mpc^{-1}}$.  This shift in the value of $H_0$ and the tightening of the constraint is the extent of the effect of the SH0ES constraint and is a typical effect when trying to calculate constraints when data sets are in tension; the center of the constraint shifts and tightens to an unreasonable degree. This reinforces the idea that even though evolving dark energy models are preferred by the combined data sets, they do not solve the $H_0$ tension with even as much freedom as in the TDE parametrization.

\section{Assessing the preference with SDSS instead of DESI BAO}
In this section, we check how much the preference for evolving dark energy depends on using the DESI or SDSS BAO dataset.

\begin{figure*}
    \centering
    \includegraphics[width=\columnwidth]{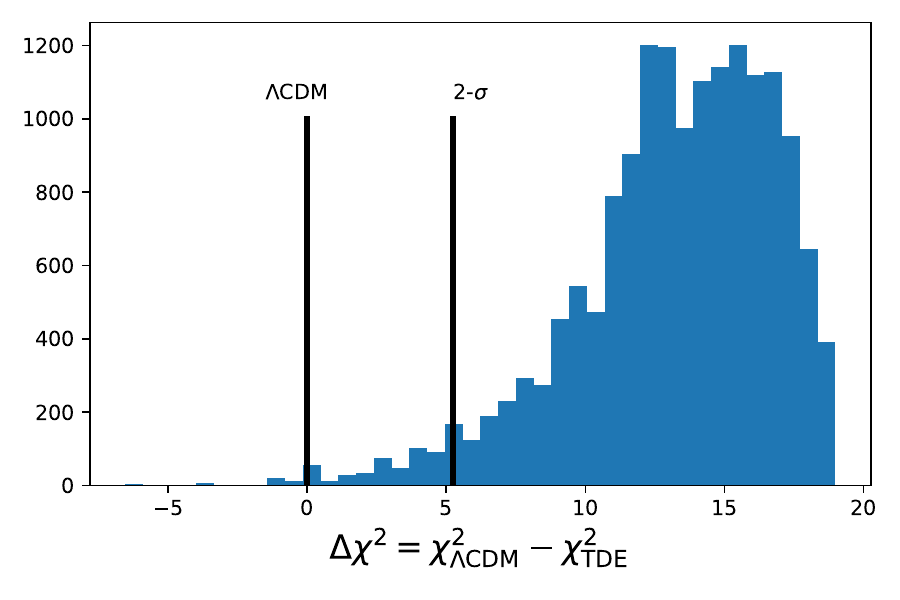}
    \includegraphics[width=\columnwidth]{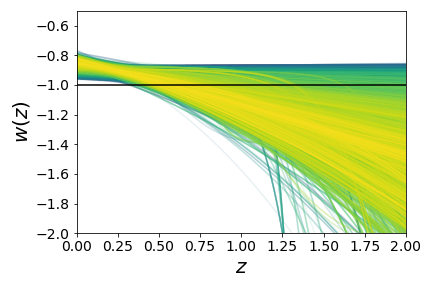}
    \caption{The $\Delta\chi^2$ distribution (left) and the $w(z)$ posterior for the case where the TDE model is constrained with CMB, SDSS BAO and DES. The $\chi^2$ values for the best-fit $\Lambda$CDM and the $2-\sigma$ TDE models are labeled.}
    \label{fig:SDSS}
\end{figure*}

In Fig.~\ref{fig:SDSS}, we see that the preference for evolving dark energy is less significant, but not by an extreme degree, down to $3\sigma$.

\section{Assessing the preference without the $z<0.1$ supernovae}
In this section, we check how much the preference for evolving dark energy depends on the $z<0.1$ SN for the DES compilation. We do this by explicitly excluding them in the likelihood and recalculating the results. In \citet{Efstathiou:2024xcq}, they showed that there is a systematic shift in the distances inferred by the DES SN and the PP SN at $z<0.1$ and hypothesized that it is these SN at $z<0.1$ that are driving the preference for evolving dark energy. 

\begin{figure*}
\centering
\includegraphics[width=\columnwidth]{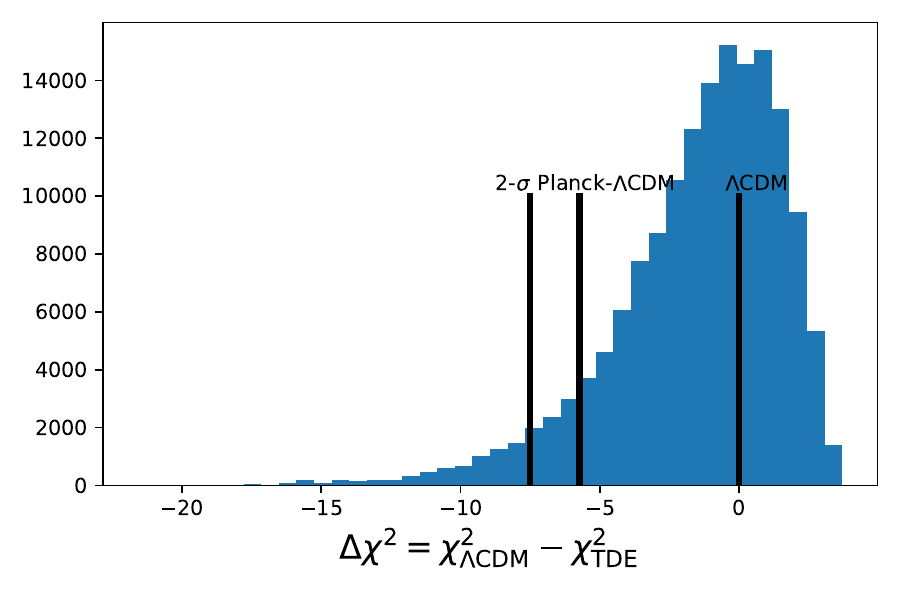}
\includegraphics[width=\columnwidth]{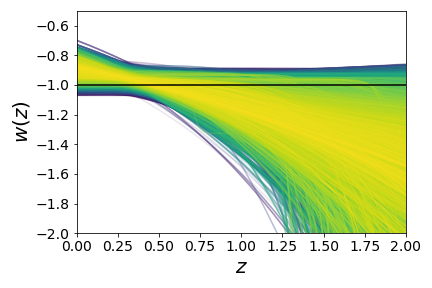}
\caption{The $\Delta\chi^2$ distribution (left) and the $w(z)$ posterior for the case where the TDE model is constrained with CMB, DESI and the low-z cut DES datasets. The $\chi^2$ values for the best-fit $\Lambda$CDM, the Planck-$\Lambda$CDM, and the $2-\sigma$ TDE models are labeled.}
\label{fig:DEScutloz}
\end{figure*}

In Fig.~\ref{fig:DEScutloz}, we see on the left the $\chi^2$ distribution for the case where the TDE parametrization is fit to the CMB, DESI and the DES dataset excluding the SN at $z<0.1$. We label both the best-fit $\Lambda$CDM model as well as the Planck-$\Lambda$CDM model.  While the data in this case are very mildly discrepant with the Planck-$\Lambda$CDM model ($<2-\sigma$) the flexibility within the $\Lambda$CDM model (varying $H_0$ and $\Omega_m$) allows a $\Delta \chi^2\sim 6$ better fit to the data in this case. This supports the idea that the the $z<0.1$ SN in the DES compilation are a key driver for the preference for the evolving dark energy that DESI finds.

We see on the right of Fig.~\ref{fig:DEScutloz} the $w(z)$ posterior for case where the TDE parametrization is fit to the CMB, DESI and the DES dataset excluding the SN at $z<0.1$. The same kinds of TDE $w(z)$ functions are still preferred over $w=-1$, just much less significantly in this case.

\begin{figure}
    \centering
    \includegraphics[width=\linewidth]{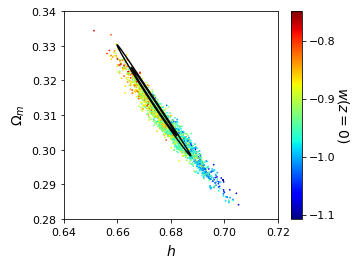}
    \caption{Posterior of $h=H_0/100$km/s/Mpc and $\Omega_m$ for the case where the TDE model is constrained by the CMB, DESI, and the DES dataset excluding the SN at $z<0.1$ (colored points) as well as the case where the $\Lambda$CDM model is constrained by just the CMB (black contours). The colored points are colored by the corresponding $w(z=0)$. }
    \label{fig:cutTension}
\end{figure}

Fig.~\ref{fig:cutTension} shows that even without the $z<0.1$ DES SN, there is a tension between the CMB and the BAO and DES SN within the $\Lambda$CDM model.  The $H_0$ and $\Omega_m$ values that the $w(z=0)=-1$ points prefer (in the context of a TDE fit to CMB, DESI, and the DES dataset excluding the SN at $z<0.1$) are different than the $H_0$ and $\Omega_m$ values the $\Lambda$CDM model prefers when constrained by just the CMB.

\section{Assessing the preference without the DESI constraint}
In this section, we check how much the preference for evolving dark energy depends on the DESI vs SN databases by calculating the constraints on $w(z)$ with CMB and 5-year DES or PP datasets.

\begin{figure*}
    \centering
    \includegraphics[width=\columnwidth]{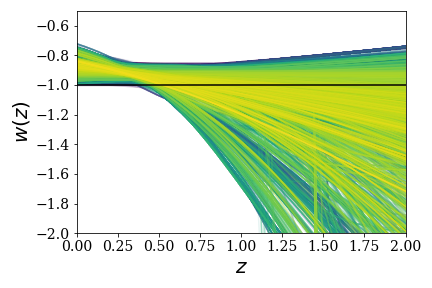}
    \includegraphics[width=\columnwidth]{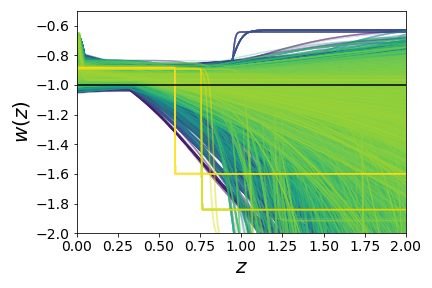}
    \caption{Like Fig.~\ref{fig:TDE_wz_post} but for the case where the TDE parametrization is constrained by the CMB+DES (left) and CMB+PP (right) joint datasets, without the inclusion of DESI BAO data.}
    \label{fig:No-DESI}
\end{figure*}

\begin{figure*}
    \centering
    \includegraphics[width=\columnwidth]{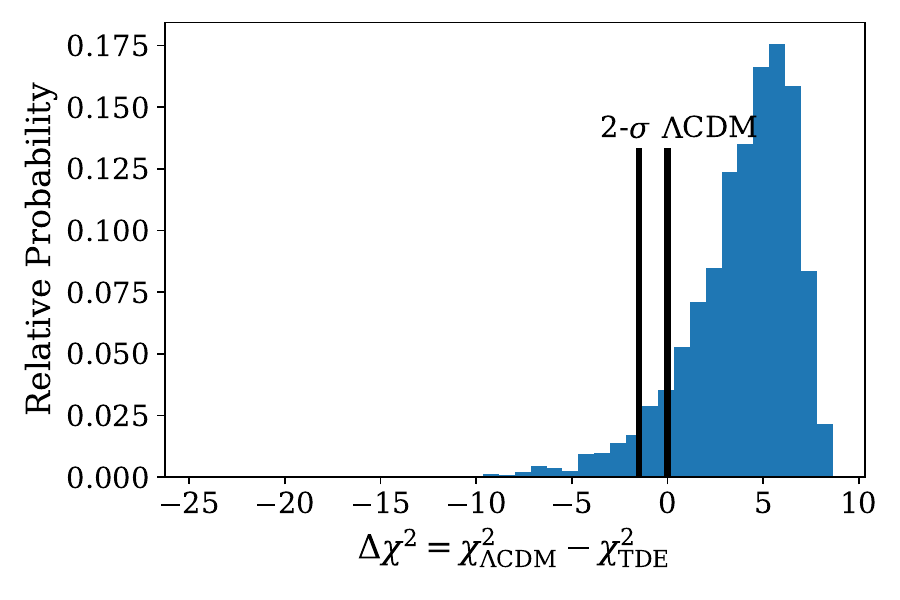}
    \includegraphics[width=\columnwidth]{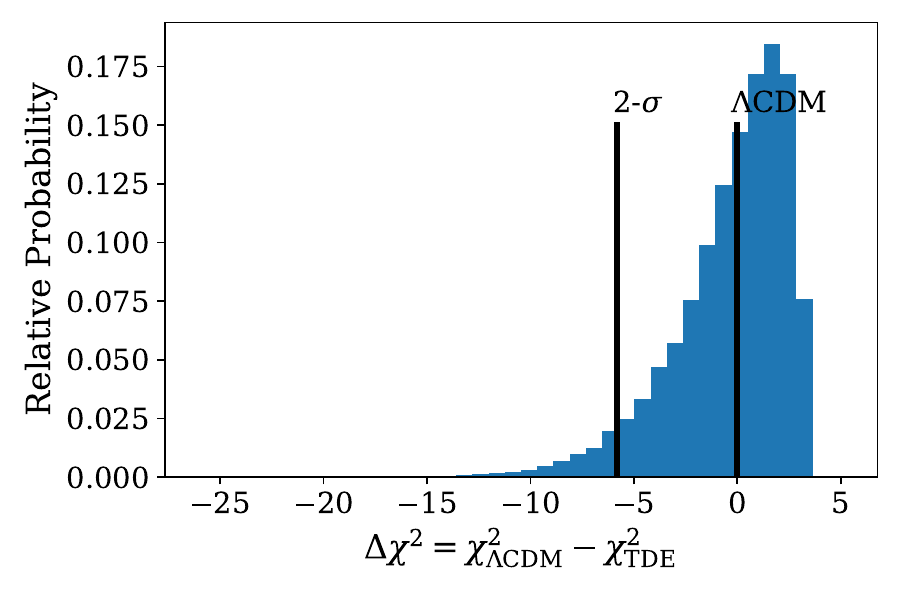}
    \caption{Like Fig.~\ref{fig:TDE_dchi2} but for the case where the TDE parametrization is constrained by the CMB+DES (left) and CMB+PP (right) joint datasets.}
    \label{fig:No-DESI-chi2}
\end{figure*}

\begin{figure*}
    \centering
    \includegraphics[width=\columnwidth]{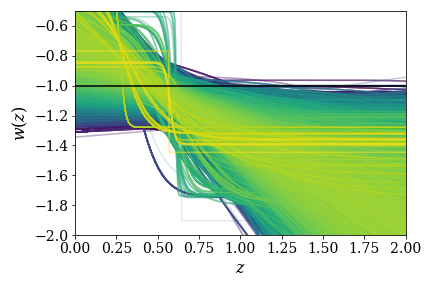}
    \includegraphics[width=\columnwidth]{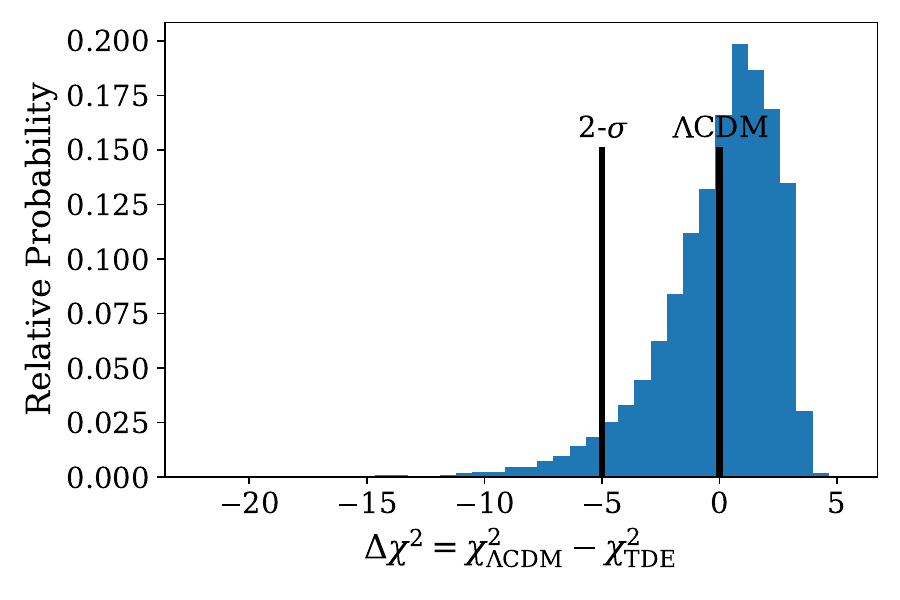}
    \caption{Like Figs.~\ref{fig:TDE_wz_post} and Fig.~\ref{fig:TDE_dchi2} but for the case where the TDE parametrization is constrained by the CMB+DESI joint dataset.}
    \label{fig:TDE_CMB_DESI}
\end{figure*}

In Figs.~\ref{fig:No-DESI} and \ref{fig:No-DESI-chi2}, we show the results for the case when we do not include the DESI result. In Fig.~\ref{fig:TDE_CMB_DESI}, we show the result for the case when we do not include either of the SN datasets.
In each of these cases, no significant deviation from $\Lambda$CDM is preferred. With only the SN or only the DESI dataset, the deviations of the DESI or SN dataset away from the best-fit Planck $\Lambda$CDM model can be explained simply by changing $H_0$ and $\Omega_m$ within $\Lambda$CDM. 

In Table~\ref{tab:chi2_table}, we present the the best-fit $\Delta \chi^2$ values ($\chi^2_{\Lambda\rm{CDM}} - \chi^2_{\rm TDE}$) as well as the reduced $\chi^2$.  One interesting aspect of the table is that for both the individual DESI and SN parts of the full CMB+DESI+SN constraint, are better fit by the TDE parametrization than the $\Lambda$CDM model.

Another interesting result of Table~\ref{tab:chi2_table} is that both PP as well as DES suffer from the problem of overly inflated covariance matrices as first pointed out in \cite{Keeley:2022iba}. It is unclear what effect this might have, as the source of the overestimated covariance matrices is not certain, but it may be the case that reduced error bars on both the DES and PP SN datasets could yield even stronger preferences for evolving dark energy.

\begin{table}
    \centering
    \begin{tabular}{c|c|c|c|c}
    \hline
    \hline
        Datasets      & DESI $\Delta \chi^2$ & $\chi^2/\rm{d.o.f}$ & SN $\Delta \chi^2$ & $\chi^2/\rm{d.o.f}$     \\
        \hline
        CMB+DESI+DES  & 12.0 & 1.04 & 17.8 & 0.923  \\
        \hline
        CMB+DESI+PP  & 7.1 & 1.07 & 12.1 & 0.930 \\
        \hline
        CMB+DES       & - & - & 6.9 & 0.922 \\
        \hline
        CMB+PP      & - & - & 3.6 & 0.930 \\
        \hline
        CMB+DESI      & 3.4 & 1.01 & - & - \\
    \end{tabular}
    \caption{best-fit $\Delta \chi^2$ and $\chi^2$ values per degree of freedom for different datasets when the TDE parametrization is fit to different combinations of datasets.}
    \label{tab:chi2_table}
\end{table}

Aside from being less significant, the PP constraint prefers slightly different regions of parameter space than the 5-year DES and so both the best-fit $\Lambda$CDM and best-fit TDE parameters will be different in the CMB+DESI+DES case compared to the CMB+DESI+PP case, and vice versa. Further, on their own, a fit to the CMB+DES, CMB+PP or CMB+DESI joint datasets drags the CMB likelihood away from the CMB best-fit while the CMB+DESI+DES or CMB+DESI+PP constraints drag the CMB likelihood to its best-fit.

From the results of this section, we see that both the BAO and the SN data are  pulling the expansion history approximately in the same direction.  Individually, the statistical preference for this deviation from $\Lambda$CDM is marginal, but when combined, their significance is intriguing.

\bibliography{main}

\begin{thebibliography}{49}%
\makeatletter
\providecommand \@ifxundefined [1]{%
 \@ifx{#1\undefined}
}%
\providecommand \@ifnum [1]{%
 \ifnum #1\expandafter \@firstoftwo
 \else \expandafter \@secondoftwo
 \fi
}%
\providecommand \@ifx [1]{%
 \ifx #1\expandafter \@firstoftwo
 \else \expandafter \@secondoftwo
 \fi
}%
\providecommand \natexlab [1]{#1}%
\providecommand \enquote  [1]{``#1''}%
\providecommand \bibnamefont  [1]{#1}%
\providecommand \bibfnamefont [1]{#1}%
\providecommand \citenamefont [1]{#1}%
\providecommand \href@noop [0]{\@secondoftwo}%
\providecommand \href [0]{\begingroup \@sanitize@url \@href}%
\providecommand \@href[1]{\@@startlink{#1}\@@href}%
\providecommand \@@href[1]{\endgroup#1\@@endlink}%
\providecommand \@sanitize@url [0]{\catcode `\\12\catcode `\$12\catcode `\&12\catcode `\#12\catcode `\^12\catcode `\_12\catcode `\%12\relax}%
\providecommand \@@startlink[1]{}%
\providecommand \@@endlink[0]{}%
\providecommand \url  [0]{\begingroup\@sanitize@url \@url }%
\providecommand \@url [1]{\endgroup\@href {#1}{\urlprefix }}%
\providecommand \urlprefix  [0]{URL }%
\providecommand \Eprint [0]{\href }%
\providecommand \doibase [0]{https://doi.org/}%
\providecommand \selectlanguage [0]{\@gobble}%
\providecommand \bibinfo  [0]{\@secondoftwo}%
\providecommand \bibfield  [0]{\@secondoftwo}%
\providecommand \translation [1]{[#1]}%
\providecommand \BibitemOpen [0]{}%
\providecommand \bibitemStop [0]{}%
\providecommand \bibitemNoStop [0]{.\EOS\space}%
\providecommand \EOS [0]{\spacefactor3000\relax}%
\providecommand \BibitemShut  [1]{\csname bibitem#1\endcsname}%
\let\auto@bib@innerbib\@empty
\bibitem [{\citenamefont {Riess}\ \emph {et~al.}(1998)\citenamefont {Riess} \emph {et~al.}}]{SupernovaSearchTeam:1998fmf}%
  \BibitemOpen
  \bibfield  {author} {\bibinfo {author} {\bibfnamefont {A.~G.}\ \bibnamefont {Riess}} \emph {et~al.} (\bibinfo {collaboration} {Supernova Search Team}),\ }\bibfield  {title} {\bibinfo {title} {{Observational evidence from supernovae for an accelerating universe and a cosmological constant}},\ }\href {https://doi.org/10.1086/300499} {\bibfield  {journal} {\bibinfo  {journal} {Astron. J.}\ }\textbf {\bibinfo {volume} {116}},\ \bibinfo {pages} {1009} (\bibinfo {year} {1998})},\ \Eprint {https://arxiv.org/abs/astro-ph/9805201} {arXiv:astro-ph/9805201} \BibitemShut {NoStop}%
\bibitem [{\citenamefont {Perlmutter}\ \emph {et~al.}(1997)\citenamefont {Perlmutter} \emph {et~al.}}]{SupernovaCosmologyProject:1997czu}%
  \BibitemOpen
  \bibfield  {author} {\bibinfo {author} {\bibfnamefont {S.}~\bibnamefont {Perlmutter}} \emph {et~al.} (\bibinfo {collaboration} {Supernova Cosmology Project}),\ }\bibfield  {title} {\bibinfo {title} {{Cosmology from Type Ia supernovae}},\ }\href@noop {} {\bibfield  {journal} {\bibinfo  {journal} {Bull. Am. Astron. Soc.}\ }\textbf {\bibinfo {volume} {29}},\ \bibinfo {pages} {1351} (\bibinfo {year} {1997})},\ \Eprint {https://arxiv.org/abs/astro-ph/9812473} {arXiv:astro-ph/9812473} \BibitemShut {NoStop}%
\bibitem [{\citenamefont {Frieman}\ \emph {et~al.}(2008)\citenamefont {Frieman}, \citenamefont {Turner},\ and\ \citenamefont {Huterer}}]{Frieman:2008sn}%
  \BibitemOpen
  \bibfield  {author} {\bibinfo {author} {\bibfnamefont {J.}~\bibnamefont {Frieman}}, \bibinfo {author} {\bibfnamefont {M.}~\bibnamefont {Turner}},\ and\ \bibinfo {author} {\bibfnamefont {D.}~\bibnamefont {Huterer}},\ }\bibfield  {title} {\bibinfo {title} {{Dark Energy and the Accelerating Universe}},\ }\href {https://doi.org/10.1146/annurev.astro.46.060407.145243} {\bibfield  {journal} {\bibinfo  {journal} {Ann. Rev. Astron. Astrophys.}\ }\textbf {\bibinfo {volume} {46}},\ \bibinfo {pages} {385} (\bibinfo {year} {2008})},\ \Eprint {https://arxiv.org/abs/0803.0982} {arXiv:0803.0982 [astro-ph]} \BibitemShut {NoStop}%
\bibitem [{\citenamefont {{Weinberg}}\ \emph {et~al.}(2013)\citenamefont {{Weinberg}}, \citenamefont {{Mortonson}}, \citenamefont {{Eisenstein}}, \citenamefont {{Hirata}}, \citenamefont {{Riess}},\ and\ \citenamefont {{Rozo}}}]{2013PhR...530...87W}%
  \BibitemOpen
  \bibfield  {author} {\bibinfo {author} {\bibfnamefont {D.~H.}\ \bibnamefont {{Weinberg}}}, \bibinfo {author} {\bibfnamefont {M.~J.}\ \bibnamefont {{Mortonson}}}, \bibinfo {author} {\bibfnamefont {D.~J.}\ \bibnamefont {{Eisenstein}}}, \bibinfo {author} {\bibfnamefont {C.}~\bibnamefont {{Hirata}}}, \bibinfo {author} {\bibfnamefont {A.~G.}\ \bibnamefont {{Riess}}},\ and\ \bibinfo {author} {\bibfnamefont {E.}~\bibnamefont {{Rozo}}},\ }\bibfield  {title} {\bibinfo {title} {{Observational probes of cosmic acceleration}},\ }\href {https://doi.org/10.1016/j.physrep.2013.05.001} {\bibfield  {journal} {\bibinfo  {journal} {Physics Reports}\ }\textbf {\bibinfo {volume} {530}},\ \bibinfo {pages} {87} (\bibinfo {year} {2013})},\ \Eprint {https://arxiv.org/abs/1201.2434} {arXiv:1201.2434 [astro-ph.CO]} \BibitemShut {NoStop}%
\bibitem [{\citenamefont {Scolnic}\ \emph {et~al.}(2018)\citenamefont {Scolnic} \emph {et~al.}}]{Pan-STARRS1:2017jku}%
  \BibitemOpen
  \bibfield  {author} {\bibinfo {author} {\bibfnamefont {D.~M.}\ \bibnamefont {Scolnic}} \emph {et~al.} (\bibinfo {collaboration} {Pan-STARRS1}),\ }\bibfield  {title} {\bibinfo {title} {{The Complete Light-curve Sample of Spectroscopically Confirmed SNe Ia from Pan-STARRS1 and Cosmological Constraints from the Combined Pantheon Sample}},\ }\href {https://doi.org/10.3847/1538-4357/aab9bb} {\bibfield  {journal} {\bibinfo  {journal} {Astrophys. J.}\ }\textbf {\bibinfo {volume} {859}},\ \bibinfo {pages} {101} (\bibinfo {year} {2018})},\ \Eprint {https://arxiv.org/abs/1710.00845} {arXiv:1710.00845 [astro-ph.CO]} \BibitemShut {NoStop}%
\bibitem [{\citenamefont {Eisenstein}\ \emph {et~al.}(2005)\citenamefont {Eisenstein} \emph {et~al.}}]{SDSS:2005xqv}%
  \BibitemOpen
  \bibfield  {author} {\bibinfo {author} {\bibfnamefont {D.~J.}\ \bibnamefont {Eisenstein}} \emph {et~al.} (\bibinfo {collaboration} {SDSS}),\ }\bibfield  {title} {\bibinfo {title} {{Detection of the Baryon Acoustic Peak in the Large-Scale Correlation Function of SDSS Luminous Red Galaxies}},\ }\href {https://doi.org/10.1086/466512} {\bibfield  {journal} {\bibinfo  {journal} {Astrophys. J.}\ }\textbf {\bibinfo {volume} {633}},\ \bibinfo {pages} {560} (\bibinfo {year} {2005})},\ \Eprint {https://arxiv.org/abs/astro-ph/0501171} {arXiv:astro-ph/0501171} \BibitemShut {NoStop}%
\bibitem [{\citenamefont {Bennett}\ \emph {et~al.}(1996)\citenamefont {Bennett}, \citenamefont {Banday}, \citenamefont {Gorski}, \citenamefont {Hinshaw}, \citenamefont {Jackson}, \citenamefont {Keegstra}, \citenamefont {Kogut}, \citenamefont {Smoot}, \citenamefont {Wilkinson},\ and\ \citenamefont {Wright}}]{Bennett:1996ce}%
  \BibitemOpen
  \bibfield  {author} {\bibinfo {author} {\bibfnamefont {C.~L.}\ \bibnamefont {Bennett}}, \bibinfo {author} {\bibfnamefont {A.}~\bibnamefont {Banday}}, \bibinfo {author} {\bibfnamefont {K.~M.}\ \bibnamefont {Gorski}}, \bibinfo {author} {\bibfnamefont {G.}~\bibnamefont {Hinshaw}}, \bibinfo {author} {\bibfnamefont {P.}~\bibnamefont {Jackson}}, \bibinfo {author} {\bibfnamefont {P.}~\bibnamefont {Keegstra}}, \bibinfo {author} {\bibfnamefont {A.}~\bibnamefont {Kogut}}, \bibinfo {author} {\bibfnamefont {G.~F.}\ \bibnamefont {Smoot}}, \bibinfo {author} {\bibfnamefont {D.~T.}\ \bibnamefont {Wilkinson}},\ and\ \bibinfo {author} {\bibfnamefont {E.~L.}\ \bibnamefont {Wright}},\ }\bibfield  {title} {\bibinfo {title} {{Four year COBE DMR cosmic microwave background observations: Maps and basic results}},\ }\href {https://doi.org/10.1086/310075} {\bibfield  {journal} {\bibinfo  {journal} {Astrophys. J. Lett.}\ }\textbf {\bibinfo {volume} {464}},\ \bibinfo {pages} {L1} (\bibinfo {year} {1996})},\ \Eprint
  {https://arxiv.org/abs/astro-ph/9601067} {arXiv:astro-ph/9601067} \BibitemShut {NoStop}%
\bibitem [{\citenamefont {Brout}\ \emph {et~al.}(2022)\citenamefont {Brout} \emph {et~al.}}]{Brout:2022vxf}%
  \BibitemOpen
  \bibfield  {author} {\bibinfo {author} {\bibfnamefont {D.}~\bibnamefont {Brout}} \emph {et~al.},\ }\bibfield  {title} {\bibinfo {title} {{The Pantheon+ Analysis: Cosmological Constraints}},\ }\href {https://doi.org/10.3847/1538-4357/ac8e04} {\bibfield  {journal} {\bibinfo  {journal} {Astrophys. J.}\ }\textbf {\bibinfo {volume} {938}},\ \bibinfo {pages} {110} (\bibinfo {year} {2022})},\ \Eprint {https://arxiv.org/abs/2202.04077} {arXiv:2202.04077 [astro-ph.CO]} \BibitemShut {NoStop}%
\bibitem [{\citenamefont {Abbott}\ \emph {et~al.}(2024)\citenamefont {Abbott} \emph {et~al.}}]{DES:2024tys}%
  \BibitemOpen
  \bibfield  {author} {\bibinfo {author} {\bibfnamefont {T.~M.~C.}\ \bibnamefont {Abbott}} \emph {et~al.} (\bibinfo {collaboration} {DES}),\ }\bibfield  {title} {\bibinfo {title} {{The Dark Energy Survey: Cosmology Results With \textasciitilde{}1500 New High-redshift Type Ia Supernovae Using The Full 5-year Dataset}},\ }\href@noop {} {\  (\bibinfo {year} {2024})},\ \Eprint {https://arxiv.org/abs/2401.02929} {arXiv:2401.02929 [astro-ph.CO]} \BibitemShut {NoStop}%
\bibitem [{\citenamefont {Rubin}\ \emph {et~al.}(2023)\citenamefont {Rubin} \emph {et~al.}}]{Rubin:2023ovl}%
  \BibitemOpen
  \bibfield  {author} {\bibinfo {author} {\bibfnamefont {D.}~\bibnamefont {Rubin}} \emph {et~al.},\ }\bibfield  {title} {\bibinfo {title} {{Union Through UNITY: Cosmology with 2,000 SNe Using a Unified Bayesian Framework}},\ }\href@noop {} {\  (\bibinfo {year} {2023})},\ \Eprint {https://arxiv.org/abs/2311.12098} {arXiv:2311.12098 [astro-ph.CO]} \BibitemShut {NoStop}%
\bibitem [{\citenamefont {Alam}\ \emph {et~al.}(2021)\citenamefont {Alam} \emph {et~al.}}]{eBOSS:2020yzd}%
  \BibitemOpen
  \bibfield  {author} {\bibinfo {author} {\bibfnamefont {S.}~\bibnamefont {Alam}} \emph {et~al.} (\bibinfo {collaboration} {eBOSS}),\ }\bibfield  {title} {\bibinfo {title} {{Completed SDSS-IV extended Baryon Oscillation Spectroscopic Survey: Cosmological implications from two decades of spectroscopic surveys at the Apache Point Observatory}},\ }\href {https://doi.org/10.1103/PhysRevD.103.083533} {\bibfield  {journal} {\bibinfo  {journal} {Phys. Rev. D}\ }\textbf {\bibinfo {volume} {103}},\ \bibinfo {pages} {083533} (\bibinfo {year} {2021})},\ \Eprint {https://arxiv.org/abs/2007.08991} {arXiv:2007.08991 [astro-ph.CO]} \BibitemShut {NoStop}%
\bibitem [{\citenamefont {Adame}\ \emph {et~al.}(2024{\natexlab{a}})\citenamefont {Adame} \emph {et~al.}}]{DESI:2024mwx}%
  \BibitemOpen
  \bibfield  {author} {\bibinfo {author} {\bibfnamefont {A.~G.}\ \bibnamefont {Adame}} \emph {et~al.} (\bibinfo {collaboration} {DESI}),\ }\bibfield  {title} {\bibinfo {title} {{DESI 2024 VI: Cosmological Constraints from the Measurements of Baryon Acoustic Oscillations}},\ }\href@noop {} {\  (\bibinfo {year} {2024}{\natexlab{a}})},\ \Eprint {https://arxiv.org/abs/2404.03002} {arXiv:2404.03002 [astro-ph.CO]} \BibitemShut {NoStop}%
\bibitem [{\citenamefont {Aghanim}\ \emph {et~al.}(2020)\citenamefont {Aghanim} \emph {et~al.}}]{Planck:2018vyg}%
  \BibitemOpen
  \bibfield  {author} {\bibinfo {author} {\bibfnamefont {N.}~\bibnamefont {Aghanim}} \emph {et~al.} (\bibinfo {collaboration} {Planck}),\ }\bibfield  {title} {\bibinfo {title} {{Planck 2018 results. VI. Cosmological parameters}},\ }\href {https://doi.org/10.1051/0004-6361/201833910} {\bibfield  {journal} {\bibinfo  {journal} {Astron. Astrophys.}\ }\textbf {\bibinfo {volume} {641}},\ \bibinfo {pages} {A6} (\bibinfo {year} {2020})},\ \bibinfo {note} {[Erratum: Astron.Astrophys. 652, C4 (2021)]},\ \Eprint {https://arxiv.org/abs/1807.06209} {arXiv:1807.06209 [astro-ph.CO]} \BibitemShut {NoStop}%
\bibitem [{\citenamefont {Qu}\ \emph {et~al.}(2024)\citenamefont {Qu} \emph {et~al.}}]{ACT:2023dou}%
  \BibitemOpen
  \bibfield  {author} {\bibinfo {author} {\bibfnamefont {F.~J.}\ \bibnamefont {Qu}} \emph {et~al.} (\bibinfo {collaboration} {ACT}),\ }\bibfield  {title} {\bibinfo {title} {{The Atacama Cosmology Telescope: A Measurement of the DR6 CMB Lensing Power Spectrum and Its Implications for Structure Growth}},\ }\href {https://doi.org/10.3847/1538-4357/acfe06} {\bibfield  {journal} {\bibinfo  {journal} {Astrophys. J.}\ }\textbf {\bibinfo {volume} {962}},\ \bibinfo {pages} {112} (\bibinfo {year} {2024})},\ \Eprint {https://arxiv.org/abs/2304.05202} {arXiv:2304.05202 [astro-ph.CO]} \BibitemShut {NoStop}%
\bibitem [{\citenamefont {Madhavacheril}\ \emph {et~al.}(2024)\citenamefont {Madhavacheril} \emph {et~al.}}]{ACT:2023kun}%
  \BibitemOpen
  \bibfield  {author} {\bibinfo {author} {\bibfnamefont {M.~S.}\ \bibnamefont {Madhavacheril}} \emph {et~al.} (\bibinfo {collaboration} {ACT}),\ }\bibfield  {title} {\bibinfo {title} {{The Atacama Cosmology Telescope: DR6 Gravitational Lensing Map and Cosmological Parameters}},\ }\href {https://doi.org/10.3847/1538-4357/acff5f} {\bibfield  {journal} {\bibinfo  {journal} {Astrophys. J.}\ }\textbf {\bibinfo {volume} {962}},\ \bibinfo {pages} {113} (\bibinfo {year} {2024})},\ \Eprint {https://arxiv.org/abs/2304.05203} {arXiv:2304.05203 [astro-ph.CO]} \BibitemShut {NoStop}%
\bibitem [{\citenamefont {Pan}\ \emph {et~al.}(2023)\citenamefont {Pan} \emph {et~al.}}]{SPT:2023jql}%
  \BibitemOpen
  \bibfield  {author} {\bibinfo {author} {\bibfnamefont {Z.}~\bibnamefont {Pan}} \emph {et~al.} (\bibinfo {collaboration} {SPT}),\ }\bibfield  {title} {\bibinfo {title} {{Measurement of gravitational lensing of the cosmic microwave background using SPT-3G 2018 data}},\ }\href {https://doi.org/10.1103/PhysRevD.108.122005} {\bibfield  {journal} {\bibinfo  {journal} {Phys. Rev. D}\ }\textbf {\bibinfo {volume} {108}},\ \bibinfo {pages} {122005} (\bibinfo {year} {2023})},\ \Eprint {https://arxiv.org/abs/2308.11608} {arXiv:2308.11608 [astro-ph.CO]} \BibitemShut {NoStop}%
\bibitem [{\citenamefont {Lodha}\ \emph {et~al.}(2025)\citenamefont {Lodha} \emph {et~al.}}]{DESI:2024kob}%
  \BibitemOpen
  \bibfield  {author} {\bibinfo {author} {\bibfnamefont {K.}~\bibnamefont {Lodha}} \emph {et~al.} (\bibinfo {collaboration} {DESI}),\ }\bibfield  {title} {\bibinfo {title} {{DESI 2024: Constraints on physics-focused aspects of dark energy using DESI DR1 BAO data}},\ }\href {https://doi.org/10.1103/PhysRevD.111.023532} {\bibfield  {journal} {\bibinfo  {journal} {Phys. Rev. D}\ }\textbf {\bibinfo {volume} {111}},\ \bibinfo {pages} {023532} (\bibinfo {year} {2025})},\ \Eprint {https://arxiv.org/abs/2405.13588} {arXiv:2405.13588 [astro-ph.CO]} \BibitemShut {NoStop}%
\bibitem [{\citenamefont {Calderon}\ \emph {et~al.}(2024)\citenamefont {Calderon} \emph {et~al.}}]{DESI:2024aqx}%
  \BibitemOpen
  \bibfield  {author} {\bibinfo {author} {\bibfnamefont {R.}~\bibnamefont {Calderon}} \emph {et~al.} (\bibinfo {collaboration} {DESI}),\ }\bibfield  {title} {\bibinfo {title} {{DESI 2024: reconstructing dark energy using crossing statistics with DESI DR1 BAO data}},\ }\href {https://doi.org/10.1088/1475-7516/2024/10/048} {\bibfield  {journal} {\bibinfo  {journal} {JCAP}\ }\textbf {\bibinfo {volume} {10}},\ \bibinfo {pages} {048}},\ \Eprint {https://arxiv.org/abs/2405.04216} {arXiv:2405.04216 [astro-ph.CO]} \BibitemShut {NoStop}%
\bibitem [{\citenamefont {Chevallier}\ and\ \citenamefont {Polarski}(2001)}]{Chevallier:2000qy}%
  \BibitemOpen
  \bibfield  {author} {\bibinfo {author} {\bibfnamefont {M.}~\bibnamefont {Chevallier}}\ and\ \bibinfo {author} {\bibfnamefont {D.}~\bibnamefont {Polarski}},\ }\bibfield  {title} {\bibinfo {title} {{Accelerating universes with scaling dark matter}},\ }\href {https://doi.org/10.1142/S0218271801000822} {\bibfield  {journal} {\bibinfo  {journal} {Int. J. Mod. Phys. D}\ }\textbf {\bibinfo {volume} {10}},\ \bibinfo {pages} {213} (\bibinfo {year} {2001})},\ \Eprint {https://arxiv.org/abs/gr-qc/0009008} {arXiv:gr-qc/0009008} \BibitemShut {NoStop}%
\bibitem [{\citenamefont {Linder}(2003)}]{Linder:2002et}%
  \BibitemOpen
  \bibfield  {author} {\bibinfo {author} {\bibfnamefont {E.~V.}\ \bibnamefont {Linder}},\ }\bibfield  {title} {\bibinfo {title} {{Exploring the expansion history of the universe}},\ }\href {https://doi.org/10.1103/PhysRevLett.90.091301} {\bibfield  {journal} {\bibinfo  {journal} {Phys. Rev. Lett.}\ }\textbf {\bibinfo {volume} {90}},\ \bibinfo {pages} {091301} (\bibinfo {year} {2003})},\ \Eprint {https://arxiv.org/abs/astro-ph/0208512} {arXiv:astro-ph/0208512} \BibitemShut {NoStop}%
\bibitem [{\citenamefont {Riess}\ \emph {et~al.}(2016)\citenamefont {Riess} \emph {et~al.}}]{Riess:2016jrr}%
  \BibitemOpen
  \bibfield  {author} {\bibinfo {author} {\bibfnamefont {A.~G.}\ \bibnamefont {Riess}} \emph {et~al.},\ }\bibfield  {title} {\bibinfo {title} {{A 2.4\% Determination of the Local Value of the Hubble Constant}},\ }\href {https://doi.org/10.3847/0004-637X/826/1/56} {\bibfield  {journal} {\bibinfo  {journal} {Astrophys. J.}\ }\textbf {\bibinfo {volume} {826}},\ \bibinfo {pages} {56} (\bibinfo {year} {2016})},\ \Eprint {https://arxiv.org/abs/1604.01424} {arXiv:1604.01424 [astro-ph.CO]} \BibitemShut {NoStop}%
\bibitem [{\citenamefont {Riess}\ \emph {et~al.}(2019)\citenamefont {Riess}, \citenamefont {Casertano}, \citenamefont {Yuan}, \citenamefont {Macri},\ and\ \citenamefont {Scolnic}}]{Riess:2019cxk}%
  \BibitemOpen
  \bibfield  {author} {\bibinfo {author} {\bibfnamefont {A.~G.}\ \bibnamefont {Riess}}, \bibinfo {author} {\bibfnamefont {S.}~\bibnamefont {Casertano}}, \bibinfo {author} {\bibfnamefont {W.}~\bibnamefont {Yuan}}, \bibinfo {author} {\bibfnamefont {L.~M.}\ \bibnamefont {Macri}},\ and\ \bibinfo {author} {\bibfnamefont {D.}~\bibnamefont {Scolnic}},\ }\bibfield  {title} {\bibinfo {title} {{Large Magellanic Cloud Cepheid Standards Provide a 1\% Foundation for the Determination of the Hubble Constant and Stronger Evidence for Physics beyond $\Lambda$CDM}},\ }\href {https://doi.org/10.3847/1538-4357/ab1422} {\bibfield  {journal} {\bibinfo  {journal} {Astrophys. J.}\ }\textbf {\bibinfo {volume} {876}},\ \bibinfo {pages} {85} (\bibinfo {year} {2019})},\ \Eprint {https://arxiv.org/abs/1903.07603} {arXiv:1903.07603 [astro-ph.CO]} \BibitemShut {NoStop}%
\bibitem [{\citenamefont {Riess}\ \emph {et~al.}(2021)\citenamefont {Riess}, \citenamefont {Casertano}, \citenamefont {Yuan}, \citenamefont {Bowers}, \citenamefont {Macri}, \citenamefont {Zinn},\ and\ \citenamefont {Scolnic}}]{Riess:2020fzl}%
  \BibitemOpen
  \bibfield  {author} {\bibinfo {author} {\bibfnamefont {A.~G.}\ \bibnamefont {Riess}}, \bibinfo {author} {\bibfnamefont {S.}~\bibnamefont {Casertano}}, \bibinfo {author} {\bibfnamefont {W.}~\bibnamefont {Yuan}}, \bibinfo {author} {\bibfnamefont {J.~B.}\ \bibnamefont {Bowers}}, \bibinfo {author} {\bibfnamefont {L.}~\bibnamefont {Macri}}, \bibinfo {author} {\bibfnamefont {J.~C.}\ \bibnamefont {Zinn}},\ and\ \bibinfo {author} {\bibfnamefont {D.}~\bibnamefont {Scolnic}},\ }\bibfield  {title} {\bibinfo {title} {{Cosmic Distances Calibrated to 1\% Precision with Gaia EDR3 Parallaxes and Hubble Space Telescope Photometry of 75 Milky Way Cepheids Confirm Tension with $\Lambda$CDM}},\ }\href {https://doi.org/10.3847/2041-8213/abdbaf} {\bibfield  {journal} {\bibinfo  {journal} {Astrophys. J. Lett.}\ }\textbf {\bibinfo {volume} {908}},\ \bibinfo {pages} {L6} (\bibinfo {year} {2021})},\ \Eprint {https://arxiv.org/abs/2012.08534} {arXiv:2012.08534 [astro-ph.CO]} \BibitemShut {NoStop}%
\bibitem [{\citenamefont {Riess}\ \emph {et~al.}(2022)\citenamefont {Riess} \emph {et~al.}}]{Riess:2021jrx}%
  \BibitemOpen
  \bibfield  {author} {\bibinfo {author} {\bibfnamefont {A.~G.}\ \bibnamefont {Riess}} \emph {et~al.},\ }\bibfield  {title} {\bibinfo {title} {{A Comprehensive Measurement of the Local Value of the Hubble Constant with 1 km s$^{−1}$ Mpc$^{−1}$ Uncertainty from the Hubble Space Telescope and the SH0ES Team}},\ }\href {https://doi.org/10.3847/2041-8213/ac5c5b} {\bibfield  {journal} {\bibinfo  {journal} {Astrophys. J. Lett.}\ }\textbf {\bibinfo {volume} {934}},\ \bibinfo {pages} {L7} (\bibinfo {year} {2022})},\ \Eprint {https://arxiv.org/abs/2112.04510} {arXiv:2112.04510 [astro-ph.CO]} \BibitemShut {NoStop}%
\bibitem [{\citenamefont {Asgari}\ \emph {et~al.}(2021)\citenamefont {Asgari} \emph {et~al.}}]{KiDS:2020suj}%
  \BibitemOpen
  \bibfield  {author} {\bibinfo {author} {\bibfnamefont {M.}~\bibnamefont {Asgari}} \emph {et~al.} (\bibinfo {collaboration} {KiDS}),\ }\bibfield  {title} {\bibinfo {title} {{KiDS-1000 Cosmology: Cosmic shear constraints and comparison between two point statistics}},\ }\href {https://doi.org/10.1051/0004-6361/202039070} {\bibfield  {journal} {\bibinfo  {journal} {Astron. Astrophys.}\ }\textbf {\bibinfo {volume} {645}},\ \bibinfo {pages} {A104} (\bibinfo {year} {2021})},\ \Eprint {https://arxiv.org/abs/2007.15633} {arXiv:2007.15633 [astro-ph.CO]} \BibitemShut {NoStop}%
\bibitem [{\citenamefont {Abbott}\ \emph {et~al.}(2022)\citenamefont {Abbott} \emph {et~al.}}]{DES:2021wwk}%
  \BibitemOpen
  \bibfield  {author} {\bibinfo {author} {\bibfnamefont {T.~M.~C.}\ \bibnamefont {Abbott}} \emph {et~al.} (\bibinfo {collaboration} {DES}),\ }\bibfield  {title} {\bibinfo {title} {{Dark Energy Survey Year 3 results: Cosmological constraints from galaxy clustering and weak lensing}},\ }\href {https://doi.org/10.1103/PhysRevD.105.023520} {\bibfield  {journal} {\bibinfo  {journal} {Phys. Rev. D}\ }\textbf {\bibinfo {volume} {105}},\ \bibinfo {pages} {023520} (\bibinfo {year} {2022})},\ \Eprint {https://arxiv.org/abs/2105.13549} {arXiv:2105.13549 [astro-ph.CO]} \BibitemShut {NoStop}%
\bibitem [{\citenamefont {Abbott}\ \emph {et~al.}(2023)\citenamefont {Abbott} \emph {et~al.}}]{Kilo-DegreeSurvey:2023gfr}%
  \BibitemOpen
  \bibfield  {author} {\bibinfo {author} {\bibfnamefont {T.~M.~C.}\ \bibnamefont {Abbott}} \emph {et~al.} (\bibinfo {collaboration} {Kilo-Degree Survey, DES}),\ }\bibfield  {title} {\bibinfo {title} {{DES Y3 + KiDS-1000: Consistent cosmology combining cosmic shear surveys}},\ }\href {https://doi.org/10.21105/astro.2305.17173} {\bibfield  {journal} {\bibinfo  {journal} {Open J. Astrophys.}\ }\textbf {\bibinfo {volume} {6}},\ \bibinfo {pages} {2305.17173} (\bibinfo {year} {2023})},\ \Eprint {https://arxiv.org/abs/2305.17173} {arXiv:2305.17173 [astro-ph.CO]} \BibitemShut {NoStop}%
\bibitem [{\citenamefont {Keeley}\ and\ \citenamefont {Shafieloo}(2023)}]{Keeley:2022ojz}%
  \BibitemOpen
  \bibfield  {author} {\bibinfo {author} {\bibfnamefont {R.~E.}\ \bibnamefont {Keeley}}\ and\ \bibinfo {author} {\bibfnamefont {A.}~\bibnamefont {Shafieloo}},\ }\bibfield  {title} {\bibinfo {title} {{Ruling Out New Physics at Low Redshift as a Solution to the H0 Tension}},\ }\href {https://doi.org/10.1103/PhysRevLett.131.111002} {\bibfield  {journal} {\bibinfo  {journal} {Phys. Rev. Lett.}\ }\textbf {\bibinfo {volume} {131}},\ \bibinfo {pages} {111002} (\bibinfo {year} {2023})},\ \Eprint {https://arxiv.org/abs/2206.08440} {arXiv:2206.08440 [astro-ph.CO]} \BibitemShut {NoStop}%
\bibitem [{\citenamefont {Cort\^es}\ and\ \citenamefont {Liddle}(2024)}]{Cortes:2024lgw}%
  \BibitemOpen
  \bibfield  {author} {\bibinfo {author} {\bibfnamefont {M.}~\bibnamefont {Cort\^es}}\ and\ \bibinfo {author} {\bibfnamefont {A.~R.}\ \bibnamefont {Liddle}},\ }\bibfield  {title} {\bibinfo {title} {{Interpreting DESI's evidence for evolving dark energy}},\ }\href@noop {} {\  (\bibinfo {year} {2024})},\ \Eprint {https://arxiv.org/abs/2404.08056} {arXiv:2404.08056 [astro-ph.CO]} \BibitemShut {NoStop}%
\bibitem [{\citenamefont {Visinelli}\ \emph {et~al.}(2019)\citenamefont {Visinelli}, \citenamefont {Vagnozzi},\ and\ \citenamefont {Danielsson}}]{Visinelli:2019qqu}%
  \BibitemOpen
  \bibfield  {author} {\bibinfo {author} {\bibfnamefont {L.}~\bibnamefont {Visinelli}}, \bibinfo {author} {\bibfnamefont {S.}~\bibnamefont {Vagnozzi}},\ and\ \bibinfo {author} {\bibfnamefont {U.}~\bibnamefont {Danielsson}},\ }\bibfield  {title} {\bibinfo {title} {{Revisiting a negative cosmological constant from low-redshift data}},\ }\href {https://doi.org/10.3390/sym11081035} {\bibfield  {journal} {\bibinfo  {journal} {Symmetry}\ }\textbf {\bibinfo {volume} {11}},\ \bibinfo {pages} {1035} (\bibinfo {year} {2019})},\ \Eprint {https://arxiv.org/abs/1907.07953} {arXiv:1907.07953 [astro-ph.CO]} \BibitemShut {NoStop}%
\bibitem [{\citenamefont {Di~Valentino}\ \emph {et~al.}(2021)\citenamefont {Di~Valentino}, \citenamefont {Melchiorri}, \citenamefont {Mena}, \citenamefont {Pan},\ and\ \citenamefont {Yang}}]{DiValentino:2020kpf}%
  \BibitemOpen
  \bibfield  {author} {\bibinfo {author} {\bibfnamefont {E.}~\bibnamefont {Di~Valentino}}, \bibinfo {author} {\bibfnamefont {A.}~\bibnamefont {Melchiorri}}, \bibinfo {author} {\bibfnamefont {O.}~\bibnamefont {Mena}}, \bibinfo {author} {\bibfnamefont {S.}~\bibnamefont {Pan}},\ and\ \bibinfo {author} {\bibfnamefont {W.}~\bibnamefont {Yang}},\ }\bibfield  {title} {\bibinfo {title} {{Interacting Dark Energy in a closed universe}},\ }\href {https://doi.org/10.1093/mnrasl/slaa207} {\bibfield  {journal} {\bibinfo  {journal} {Mon. Not. Roy. Astron. Soc.}\ }\textbf {\bibinfo {volume} {502}},\ \bibinfo {pages} {L23} (\bibinfo {year} {2021})},\ \Eprint {https://arxiv.org/abs/2011.00283} {arXiv:2011.00283 [astro-ph.CO]} \BibitemShut {NoStop}%
\bibitem [{\citenamefont {Calder\'on}\ \emph {et~al.}(2021)\citenamefont {Calder\'on}, \citenamefont {Gannouji}, \citenamefont {L'Huillier},\ and\ \citenamefont {Polarski}}]{Calderon:2020hoc}%
  \BibitemOpen
  \bibfield  {author} {\bibinfo {author} {\bibfnamefont {R.}~\bibnamefont {Calder\'on}}, \bibinfo {author} {\bibfnamefont {R.}~\bibnamefont {Gannouji}}, \bibinfo {author} {\bibfnamefont {B.}~\bibnamefont {L'Huillier}},\ and\ \bibinfo {author} {\bibfnamefont {D.}~\bibnamefont {Polarski}},\ }\bibfield  {title} {\bibinfo {title} {{Negative cosmological constant in the dark sector?}},\ }\href {https://doi.org/10.1103/PhysRevD.103.023526} {\bibfield  {journal} {\bibinfo  {journal} {Phys. Rev. D}\ }\textbf {\bibinfo {volume} {103}},\ \bibinfo {pages} {023526} (\bibinfo {year} {2021})},\ \Eprint {https://arxiv.org/abs/2008.10237} {arXiv:2008.10237 [astro-ph.CO]} \BibitemShut {NoStop}%
\bibitem [{\citenamefont {Shlivko}\ and\ \citenamefont {Steinhardt}(2024)}]{Shlivko:2024llw}%
  \BibitemOpen
  \bibfield  {author} {\bibinfo {author} {\bibfnamefont {D.}~\bibnamefont {Shlivko}}\ and\ \bibinfo {author} {\bibfnamefont {P.~J.}\ \bibnamefont {Steinhardt}},\ }\bibfield  {title} {\bibinfo {title} {{Assessing observational constraints on dark energy}},\ }\href {https://doi.org/10.1016/j.physletb.2024.138826} {\bibfield  {journal} {\bibinfo  {journal} {Phys. Lett. B}\ }\textbf {\bibinfo {volume} {855}},\ \bibinfo {pages} {138826} (\bibinfo {year} {2024})},\ \Eprint {https://arxiv.org/abs/2405.03933} {arXiv:2405.03933 [astro-ph.CO]} \BibitemShut {NoStop}%
\bibitem [{\citenamefont {Wolf}\ \emph {et~al.}(2025)\citenamefont {Wolf}, \citenamefont {Garc\'\i{}a-Garc\'\i{}a},\ and\ \citenamefont {Ferreira}}]{Wolf:2025jlc}%
  \BibitemOpen
  \bibfield  {author} {\bibinfo {author} {\bibfnamefont {W.~J.}\ \bibnamefont {Wolf}}, \bibinfo {author} {\bibfnamefont {C.}~\bibnamefont {Garc\'\i{}a-Garc\'\i{}a}},\ and\ \bibinfo {author} {\bibfnamefont {P.~G.}\ \bibnamefont {Ferreira}},\ }\bibfield  {title} {\bibinfo {title} {{Robustness of Dark Energy Phenomenology Across Different Parameterizations}},\ }\href@noop {} {\  (\bibinfo {year} {2025})},\ \Eprint {https://arxiv.org/abs/2502.04929} {arXiv:2502.04929 [astro-ph.CO]} \BibitemShut {NoStop}%
\bibitem [{\citenamefont {Alcock}\ and\ \citenamefont {Paczynski}(1979)}]{Alcock:1979mp}%
  \BibitemOpen
  \bibfield  {author} {\bibinfo {author} {\bibfnamefont {C.}~\bibnamefont {Alcock}}\ and\ \bibinfo {author} {\bibfnamefont {B.}~\bibnamefont {Paczynski}},\ }\bibfield  {title} {\bibinfo {title} {{An evolution free test for non-zero cosmological constant}},\ }\href {https://doi.org/10.1038/281358a0} {\bibfield  {journal} {\bibinfo  {journal} {Nature}\ }\textbf {\bibinfo {volume} {281}},\ \bibinfo {pages} {358} (\bibinfo {year} {1979})}\BibitemShut {NoStop}%
\bibitem [{\citenamefont {Adame}\ \emph {et~al.}(2024{\natexlab{b}})\citenamefont {Adame} \emph {et~al.}}]{DESI:2024hhd}%
  \BibitemOpen
  \bibfield  {author} {\bibinfo {author} {\bibfnamefont {A.~G.}\ \bibnamefont {Adame}} \emph {et~al.} (\bibinfo {collaboration} {DESI}),\ }\bibfield  {title} {\bibinfo {title} {{DESI 2024 VII: Cosmological Constraints from the Full-Shape Modeling of Clustering Measurements}},\ }\href@noop {} {\  (\bibinfo {year} {2024}{\natexlab{b}})},\ \Eprint {https://arxiv.org/abs/2411.12022} {arXiv:2411.12022 [astro-ph.CO]} \BibitemShut {NoStop}%
\bibitem [{\citenamefont {Adame}\ \emph {et~al.}(2024{\natexlab{c}})\citenamefont {Adame} \emph {et~al.}}]{DESI:2024jis}%
  \BibitemOpen
  \bibfield  {author} {\bibinfo {author} {\bibfnamefont {A.~G.}\ \bibnamefont {Adame}} \emph {et~al.} (\bibinfo {collaboration} {DESI}),\ }\bibfield  {title} {\bibinfo {title} {{DESI 2024 V: Full-Shape Galaxy Clustering from Galaxies and Quasars}},\ }\href@noop {} {\  (\bibinfo {year} {2024}{\natexlab{c}})},\ \Eprint {https://arxiv.org/abs/2411.12021} {arXiv:2411.12021 [astro-ph.CO]} \BibitemShut {NoStop}%
\bibitem [{\citenamefont {Keeley}\ \emph {et~al.}(2019)\citenamefont {Keeley}, \citenamefont {Joudaki}, \citenamefont {Kaplinghat},\ and\ \citenamefont {Kirkby}}]{Keeley:2019esp}%
  \BibitemOpen
  \bibfield  {author} {\bibinfo {author} {\bibfnamefont {R.~E.}\ \bibnamefont {Keeley}}, \bibinfo {author} {\bibfnamefont {S.}~\bibnamefont {Joudaki}}, \bibinfo {author} {\bibfnamefont {M.}~\bibnamefont {Kaplinghat}},\ and\ \bibinfo {author} {\bibfnamefont {D.}~\bibnamefont {Kirkby}},\ }\bibfield  {title} {\bibinfo {title} {{Implications of a transition in the dark energy equation of state for the $H_0$ and $\sigma_8$ tensions}},\ }\href {https://doi.org/10.1088/1475-7516/2019/12/035} {\bibfield  {journal} {\bibinfo  {journal} {JCAP}\ }\textbf {\bibinfo {volume} {12}},\ \bibinfo {pages} {035}},\ \Eprint {https://arxiv.org/abs/1905.10198} {arXiv:1905.10198 [astro-ph.CO]} \BibitemShut {NoStop}%
\bibitem [{\citenamefont {Keeley}\ \emph {et~al.}(2020)\citenamefont {Keeley}, \citenamefont {Shafieloo}, \citenamefont {Hazra},\ and\ \citenamefont {Souradeep}}]{Keeley:2020rmo}%
  \BibitemOpen
  \bibfield  {author} {\bibinfo {author} {\bibfnamefont {R.~E.}\ \bibnamefont {Keeley}}, \bibinfo {author} {\bibfnamefont {A.}~\bibnamefont {Shafieloo}}, \bibinfo {author} {\bibfnamefont {D.~K.}\ \bibnamefont {Hazra}},\ and\ \bibinfo {author} {\bibfnamefont {T.}~\bibnamefont {Souradeep}},\ }\bibfield  {title} {\bibinfo {title} {{Inflation Wars: A New Hope}},\ }\href {https://doi.org/10.1088/1475-7516/2020/09/055} {\bibfield  {journal} {\bibinfo  {journal} {JCAP}\ }\textbf {\bibinfo {volume} {09}},\ \bibinfo {pages} {055}},\ \Eprint {https://arxiv.org/abs/2006.12710} {arXiv:2006.12710 [astro-ph.CO]} \BibitemShut {NoStop}%
\bibitem [{\citenamefont {{Matthewson}}\ and\ \citenamefont {{Shafieloo}}(2025)}]{2025JCAP...01..064M}%
  \BibitemOpen
  \bibfield  {author} {\bibinfo {author} {\bibfnamefont {W.~L.}\ \bibnamefont {{Matthewson}}}\ and\ \bibinfo {author} {\bibfnamefont {A.}~\bibnamefont {{Shafieloo}}},\ }\bibfield  {title} {\bibinfo {title} {{Star-crossed labours: checking consistency between current supernovae compilations}},\ }\href {https://doi.org/10.1088/1475-7516/2025/01/064} {\bibfield  {journal} {\bibinfo  {journal} {\jcap}\ }\textbf {\bibinfo {volume} {2025}},\ \bibinfo {eid} {064} (\bibinfo {year} {2025})},\ \Eprint {https://arxiv.org/abs/2409.02550} {arXiv:2409.02550 [astro-ph.CO]} \BibitemShut {NoStop}%
\bibitem [{\citenamefont {Linder}(2007)}]{Linder:2007ka}%
  \BibitemOpen
  \bibfield  {author} {\bibinfo {author} {\bibfnamefont {E.~V.}\ \bibnamefont {Linder}},\ }\bibfield  {title} {\bibinfo {title} {{The Mirage of w=-1}},\ }\href@noop {} {\  (\bibinfo {year} {2007})},\ \Eprint {https://arxiv.org/abs/0708.0024} {arXiv:0708.0024 [astro-ph]} \BibitemShut {NoStop}%
\bibitem [{\citenamefont {Dodelson}\ and\ \citenamefont {Schmidt}(2020)}]{Dodelson:2020bqr}%
  \BibitemOpen
  \bibfield  {author} {\bibinfo {author} {\bibfnamefont {S.}~\bibnamefont {Dodelson}}\ and\ \bibinfo {author} {\bibfnamefont {F.}~\bibnamefont {Schmidt}},\ }\href {https://doi.org/10.1016/C2017-0-01943-2} {\emph {\bibinfo {title} {{Modern Cosmology}}}}\ (\bibinfo  {publisher} {Academic Press},\ \bibinfo {year} {2020})\BibitemShut {NoStop}%
\bibitem [{\citenamefont {Aghamousa}\ \emph {et~al.}(2016)\citenamefont {Aghamousa} \emph {et~al.}}]{DESI:2016fyo}%
  \BibitemOpen
  \bibfield  {author} {\bibinfo {author} {\bibfnamefont {A.}~\bibnamefont {Aghamousa}} \emph {et~al.} (\bibinfo {collaboration} {DESI}),\ }\bibfield  {title} {\bibinfo {title} {{The DESI Experiment Part I: Science,Targeting, and Survey Design}},\ }\href@noop {} {\  (\bibinfo {year} {2016})},\ \Eprint {https://arxiv.org/abs/1611.00036} {arXiv:1611.00036 [astro-ph.IM]} \BibitemShut {NoStop}%
\bibitem [{\citenamefont {Hildebrandt}\ \emph {et~al.}(2017)\citenamefont {Hildebrandt} \emph {et~al.}}]{Hildebrandt:2016iqg}%
  \BibitemOpen
  \bibfield  {author} {\bibinfo {author} {\bibfnamefont {H.}~\bibnamefont {Hildebrandt}} \emph {et~al.},\ }\bibfield  {title} {\bibinfo {title} {{KiDS-450: Cosmological parameter constraints from tomographic weak gravitational lensing}},\ }\href {https://doi.org/10.1093/mnras/stw2805} {\bibfield  {journal} {\bibinfo  {journal} {Mon. Not. Roy. Astron. Soc.}\ }\textbf {\bibinfo {volume} {465}},\ \bibinfo {pages} {1454} (\bibinfo {year} {2017})},\ \Eprint {https://arxiv.org/abs/1606.05338} {arXiv:1606.05338 [astro-ph.CO]} \BibitemShut {NoStop}%
\bibitem [{\citenamefont {Nguyen}\ \emph {et~al.}(2023)\citenamefont {Nguyen}, \citenamefont {Huterer},\ and\ \citenamefont {Wen}}]{Nguyen:2023fip}%
  \BibitemOpen
  \bibfield  {author} {\bibinfo {author} {\bibfnamefont {N.-M.}\ \bibnamefont {Nguyen}}, \bibinfo {author} {\bibfnamefont {D.}~\bibnamefont {Huterer}},\ and\ \bibinfo {author} {\bibfnamefont {Y.}~\bibnamefont {Wen}},\ }\bibfield  {title} {\bibinfo {title} {{Evidence for Suppression of Structure Growth in the Concordance Cosmological Model}},\ }\href {https://doi.org/10.1103/PhysRevLett.131.111001} {\bibfield  {journal} {\bibinfo  {journal} {Phys. Rev. Lett.}\ }\textbf {\bibinfo {volume} {131}},\ \bibinfo {pages} {111001} (\bibinfo {year} {2023})},\ \Eprint {https://arxiv.org/abs/2302.01331} {arXiv:2302.01331 [astro-ph.CO]} \BibitemShut {NoStop}%
\bibitem [{\citenamefont {Keeley}\ \emph {et~al.}(2022)\citenamefont {Keeley}, \citenamefont {Shafieloo},\ and\ \citenamefont {L'Huillier}}]{Keeley:2022iba}%
  \BibitemOpen
  \bibfield  {author} {\bibinfo {author} {\bibfnamefont {R.}~\bibnamefont {Keeley}}, \bibinfo {author} {\bibfnamefont {A.}~\bibnamefont {Shafieloo}},\ and\ \bibinfo {author} {\bibfnamefont {B.}~\bibnamefont {L'Huillier}},\ }\bibfield  {title} {\bibinfo {title} {{An Analysis of Variance of the Pantheon+ Dataset: Systematics in the Covariance Matrix?}},\ }\href@noop {} {\  (\bibinfo {year} {2022})},\ \Eprint {https://arxiv.org/abs/2212.07917} {arXiv:2212.07917 [astro-ph.CO]} \BibitemShut {NoStop}%
\bibitem [{\citenamefont {Efstathiou}(2024)}]{Efstathiou:2024xcq}%
  \BibitemOpen
  \bibfield  {author} {\bibinfo {author} {\bibfnamefont {G.}~\bibnamefont {Efstathiou}},\ }\bibfield  {title} {\bibinfo {title} {{Evolving Dark Energy or Supernovae Systematics?}},\ }\href@noop {} {\  (\bibinfo {year} {2024})},\ \Eprint {https://arxiv.org/abs/2408.07175} {arXiv:2408.07175 [astro-ph.CO]} \BibitemShut {NoStop}%
\bibitem [{\citenamefont {Dinda}(2024)}]{Dinda:2024kjf}%
  \BibitemOpen
  \bibfield  {author} {\bibinfo {author} {\bibfnamefont {B.~R.}\ \bibnamefont {Dinda}},\ }\bibfield  {title} {\bibinfo {title} {{A new diagnostic for the null test of dynamical dark energy in light of DESI 2024 and other BAO data}},\ }\href {https://doi.org/10.1088/1475-7516/2024/09/062} {\bibfield  {journal} {\bibinfo  {journal} {JCAP}\ }\textbf {\bibinfo {volume} {09}},\ \bibinfo {pages} {062}},\ \Eprint {https://arxiv.org/abs/2405.06618} {arXiv:2405.06618 [astro-ph.CO]} \BibitemShut {NoStop}%
\bibitem [{\citenamefont {Efstathiou}(2021)}]{Efstathiou:2021ocp}%
  \BibitemOpen
  \bibfield  {author} {\bibinfo {author} {\bibfnamefont {G.}~\bibnamefont {Efstathiou}},\ }\bibfield  {title} {\bibinfo {title} {{To H0 or not to H0?}},\ }\href {https://doi.org/10.1093/mnras/stab1588} {\bibfield  {journal} {\bibinfo  {journal} {Mon. Not. Roy. Astron. Soc.}\ }\textbf {\bibinfo {volume} {505}},\ \bibinfo {pages} {3866} (\bibinfo {year} {2021})},\ \Eprint {https://arxiv.org/abs/2103.08723} {arXiv:2103.08723 [astro-ph.CO]} \BibitemShut {NoStop}%
\end{thebibliography}%

\end{document}